\documentclass[journal,article,submit,moreauthors]{Definitions/mdpi}
\renewcommand{\linenumbers}{}
\newcommand{\dd}{\mathrm{d}}
\firstpage{1} 
\pubvolume{1}
\issuenum{1}
\articlenumber{0}
\pubyear{2026}
\copyrightyear{2026}
\datereceived{ } 
\daterevised{ } 
\dateaccepted{ } 
\datepublished{ } 

\Title{Momentum-Space Path Integral Approach to Non-Hermitian Symmetry Breaking}

\Author{Zhuo-Ting Cai $^{1}$ and Wei Chen $^{1,2,}$*\orcidA{}}

\AuthorNames{Zhuo-Ting Cai and Wei Chen}

\address{%
$^{1}$ \quad National Laboratory of Solid State Microstructures,
School of Physics, Jiangsu Physical Science Research Center,
and Collaborative Innovation Center of Advanced Microstructures,
Nanjing University, Nanjing 210093, China;\\
$^{2}$\quad Jiangsu Physical Science Research Center and Jiangsu Key Laboratory of Quantum Information Science and Technology, Nanjing University, Nanjing 210093, China;}

\corres{Correspondence: pchenweis@gmail.com }

\abstract{A quantum-classical correspondence for non-Hermitian symmetry breaking has recently been established using coordinate-space path integrals, providing a semiclassical understanding of spectral transitions at the level of individual eigenstates. Here we develop its dual formulation in momentum space by constructing the corresponding trace formula and quantization condition. We show that the real or complex nature of individual eigenvalues is determined by the symmetry properties of the associated semiclassical orbits, as in the coordinate-space path-integral approach. 
{Moreover, we demonstrate that the topology of semiclassical orbits determines the natural formulation of the quantization condition: the coordinate- and momentum-space formulations are equivalent for contractible periodic orbits in phase space, whereas for noncontractible orbits, the quantization condition along the winding direction remains valid, but its dual form must be corrected by a boundary term.
In particular, real-space-winding and Brillouin-zone-winding orbits naturally select coordinate- and momentum-space quantization, respectively.}
As a nontrivial application, we investigate the boundary-induced spectral transition of Bloch oscillations in a finite non-Hermitian lattice, where Bloch-oscillation orbits winding across the Brillouin zone preserve the relevant symmetry and yield real energy levels, whereas boundary-reflected orbits form symmetry-related pairs and give rise to complex-conjugate eigenvalues. Our work completes the quantum-classical correspondence framework for non-Hermitian symmetry breaking, extending its applicability to a broader class of non-Hermitian problems.}


\begin{document}
\nolinenumbers
\section{Introduction}
Conventional quantum mechanics is built on Hermitian Hamiltonians, which ensure
real spectra and unitary time evolution. However, many physical systems are not perfectly isolated: they exchange particles or energy with their environments and exhibit gain or dissipation effects. 
In many cases, non-Hermitian effective Hamiltonians provide a natural description of these phenomena and have been widely researched in optics~\cite{feng2013nm,feng2017non,Malzard2015prl}, acoustics~\cite{Huang2024AcousticResonances,Zhang2021AcousticNHSE,Zhu2018prl}, and condensed-matter physics~\cite{LonghiS2019prl1,LonghiS2019prl2,chenshu2019prb,Ananya2019jop,shaokai2024prl,genghao2023prb}.
The studies have revealed a variety of intriguing phenomena, including exceptional points~\cite{Miri2019science,Dembowski2004pre,
Bergholtz2021rmp,chen2017exceptional}, non-Hermitian skin effects~\cite{yangzhesen2020AGBZ,Kunst2018prl,Yokomizo2019prl,Borgnia20prl,Zhangkai2020prl,Okuma2020prl},
unconventional topological phases~\cite{Wangzhong2018prl,XuePeng2022prl,Palacios2021nc,Kawabata2020PRR}, {and momentum-gapped dispersions~\cite{Baggioli2020Physics_Reports,Yerin2023prb}.}
Many of these phenomena are closely related to the spectral properties of non-Hermitian Hamiltonians~\cite{Kawabata2019prx,fuliang2018prl,Jazaeri2001PRE}, which are central to non-Hermitian physics.

Among various concepts in non-Hermitian physics, pseudo-Hermiticity~\cite{mostafazadeh2002pseudo,Bender98prl,Bender2007making} has attracted considerable attention. It imposes a symmetry constraint on non-Hermitian Hamiltonians that can give rise to entirely real energy spectra.
More precisely, eigenenergies are real for eigenstates that are invariant under the pseudo-Hermiticity operation, whereas eigenstates that break this symmetry are associated with complex-conjugate eigenvalue pairs.
While the mathematical foundations of pseudo-Hermiticity are well established~\cite{Ashida2020NHphysic,mostafazadeh2010pseudo} and spontaneous breaking of pseudo-Hermiticity symmetry has been experimentally observed~\cite{Guo2009prl,Ruter2010np,Regensburger2012Nature,Peng2014NP,LiangFeng2014science,ozdemir2019parity}, 
a universal mechanism governing these phenomena at the level of individual eigenstates across both symmetry-preserving and symmetry-broken regimes had remained elusive for a long time.

In a recent work~\cite{cai2025prl}, we addressed this problem by establishing a quantum-classical correspondence principle for non-Hermitian symmetry breaking. By generalizing the Gutzwiller trace formula and the associated quantization condition to the complex domain via coordinate-space path integrals, we revealed a correspondence between the energy spectrum and complex classical orbits. This correspondence allows the real or complex character of individual eigenenergies to be understood from the symmetry properties of the underlying semiclassical orbits. Specifically, real eigenenergies are associated with self-symmetric orbits, whereas complex-conjugate eigenvalue pairs originate from symmetry-related pairs of orbits. Given that our previous work was formulated in terms of coordinate-space path integrals, a natural question arises: whether an analogous quantum-classical correspondence can be established via momentum-space path integrals, and whether the two formulations provide equivalent descriptions for generic non-Hermitian systems.

In this work, we establish the quantum-classical correspondence principle for non-Hermitian symmetry breaking using a momentum-space path-integral approach. We generalize the trace formula and quantization condition to momentum space and interpret whether individual eigenvalues are real or complex in terms of the symmetry properties of their corresponding semiclassical orbits. By comparing the quantization conditions formulated in two dual representations, we clarify their respective regimes of validity. For contractible periodic orbits without phase-space winding, the two formulations are equivalent. 
{For noncontractible winding orbits, the winding direction determines which formulation is globally well defined in its conventional form: real-space winding is naturally described in coordinate space, whereas Brillouin-zone winding is naturally described in momentum space. In either case, the dual formulation requires the corresponding boundary term to recover the same quantization condition.}
As a nontrivial example where the momentum-space formulation is essential, we investigate the boundary-induced real-to-complex spectral transition of Bloch oscillations in a finite non-Hermitian lattice. The semiclassical orbits traversing the Brillouin zone preserve the corresponding symmetry and yield real energies after quantization, whereas boundary-reflected orbits break this symmetry and form symmetry-related pairs, resulting in complex-conjugate energy pairs. The spectra predicted by our theory agree well with exact diagonalization results, confirming the validity of this framework. Our work provides a comprehensive semiclassical framework for non-Hermitian symmetry breaking and extends its applicability to a broader class of systems.

\section{Quantum-classical correspondence from momentum-space path integral}
In this section, we establish the quantum-classical correspondence of non-Hermitian symmetry breaking through a momentum-space path integral, instead of the previous coordinate-space approach~\cite{cai2025prl}.
The main steps of our analysis are as follows:
First, we develop the
momentum-space path-integral framework and establish the correspondence between
the quantum $\eta$-pseudo-Hermiticity and the associated classical $S_\eta$ symmetry.
Next, we derive the extended non-Hermitian Gutzwiller trace formula in momentum space and the corresponding quantization condition. We will compare it with its coordinate-space counterpart, clarifying their relation and respective domains of applicability.
Furthermore, we relate the symmetry of semiclassical orbits to the real-or-complex nature of the quantized energy levels.


As a starting point, we consider a generic non-Hermitian system governed by the Hamiltonian $\mathcal{H}(\hat{x}, \hat{p})$. Here, $\hat{x}$ and $\hat{p}$ denote the canonical position and momentum operators, respectively. In this work, we restrict our discussion to systems that satisfy the condition of $\eta$-pseudo-Hermiticity~\cite{mostafazadeh2002pseudo}, which we refer to as $\eta$-pseudo-Hermitian symmetry ($\eta$-PHS). This fundamental symmetry requires the existence of a Hermitian, invertible operator $\eta$ (i.e., $\eta = \eta^\dagger$), such that the Hamiltonian satisfies
\begin{equation}\label{hs}
\eta \mathcal{H} \eta^{-1}=\mathcal{H}^{\dagger}.
\end{equation}

Following the same principle as the coordinate-space construction~\cite{cai2025prl}, we begin with the propagator in the momentum representation,
\begin{equation}
G(p_f,p_i;t)=\langle p_f|e^{-i\mathcal Ht}|p_i\rangle .
\end{equation}
For the subsequent analysis, we adopt a biorthogonal momentum basis and analytically continue the semiclassical phase-space variables to the complex domain.
The action of $\eta$ on the coordinate and momentum operators is denoted by
$
\hat x_\eta=\eta\hat x\eta^{-1},
\hat p_\eta=\eta\hat p\eta^{-1},
$
and its corresponding action on the propagator is obtained through
$
G_\eta(p_f,p_i;t)
=
\langle p_f|
e^{-i\mathcal H(\hat x_\eta,\hat p_\eta)t}
|p_i\rangle,
$ 
with its physical interpretation specified in concrete examples.
As shown in Appendix~\ref{App.A1}, the $\eta$-PHS~\eqref{hs} imposes the following constraint on the propagator
\begin{equation}\label{kgp}
G_\eta(p_f,p_i;t)
=
G^*(p_i^*,p_f^*;-t^*) .
\end{equation}

We next express the propagator in terms of the momentum-space path integral, which takes the form
\begin{equation}\label{eq4}
G(p_f,p_i;t)
=
\int_{p_i}^{p_f}\mathcal D[\xi]\,
\exp\left\{
i\int_0^t
\left[-x\dot p-H(x,p)\right]\mathrm{d}t'
\right\},
\end{equation}
where $\xi(t')=\{x(t'),p(t')\}$ denotes a phase-space orbit satisfying $p(0)=p_i$ and $p(t)=p_f$.
In this representation, Eq.~\eqref{kgp} becomes
\begin{equation}\label{pathp}
\begin{aligned}
&\int_{p_i}^{p_f}\mathcal D[\xi]\,
e^{i\int_0^t[-x\dot p-H(x_\eta,p_\eta)]\mathrm{d}t'}
\\=&
\int_{p_i}^{p_f}\mathcal D[\xi]\,
e^{i\int_0^t[-x\dot p-H^*(x^*,p^*)]\mathrm{d}t'},
\end{aligned}
\end{equation}
where $x_\eta=x_\eta(x,p)$ and $p_\eta=p_\eta(x,p)$ denote the classical
variables induced by the $\eta$ transformation.
Here, $x$ and $p$ are $c$ numbers which are analytically
continued to complex phase space, and $H(x,p)$ denotes the corresponding classical Hamiltonian.
Since Eq.~\eqref{pathp} holds for arbitrary $p_i$, $p_f$, and $t$, the semiclassical Hamiltonian is required to satisfy the following constraint, referred to as the $S_\eta$ symmetry:
\begin{equation}\label{sym_Hcl_p}
S_\eta:
H(x_\eta,p_\eta)=H^*(x^*,p^*),
\end{equation}
as shown in Appendix~\ref{App.A2}.
The symmetry agrees with the result obtained through the coordinate-space path integral approach~\cite{cai2025prl}.

Having established the correspondence between the quantum
$\eta$-PHS and the classical $S_\eta$ symmetry, we next relate the semiclassical orbits to the quantum energy spectrum. 
This is achieved through the trace formula in momentum space and the resulting quantization condition.
{In deriving the trace formula and throughout the subsequent discussion, we restrict our analysis to one-dimensional systems. A higher-dimensional momentum-space trace formula can, in principle,
be constructed in analogy with the multidimensional coordinate-space
trace formula~\cite{gutzwiller1971JMP}.}
We begin with the trace of the Green operator, \(G(E)=\mathrm{Tr}(\mathcal H-E)^{-1}\), whose poles determine the eigenenergies.
Using the biorthogonal momentum basis, it can be written as
\begin{equation}
G(E)
=
i\int \mathrm{d}t\, e^{iEt}
\int \mathrm{d}p_i\,
\langle p_i|e^{-i\mathcal Ht}|p_i\rangle .
\end{equation}
Substituting Eq.~\eqref{eq4}, we obtain the momentum-space path-integral representation
\begin{equation}\label{Gp_path}
G(E)
=
i\int \mathrm{d}t\, e^{iEt}
\int \mathrm{d}p_i
\int_{p_i}^{p_i}\mathcal D[\xi]\,
e^{
i\int_0^t
\left[-x\dot p-H(x,p)\right]\mathrm{d}t'
}.
\end{equation}
The boundary condition $p(0)=p(t)=p_i$ imposed by the trace selects closed orbits in momentum space. For a periodic Brillouin zone, however, closed orbits are defined modulo the reciprocal-lattice period $2\pi$,
such that $p(t)=p(0)+2\pi w$, where $w\in\mathbb Z$ is the winding number.

Equation~\eqref{Gp_path} represents a path integral over all phase-space orbits $\xi$ starting and ending at $p_i$ during a propagation time $t$.
We evaluate these three integrations sequentially using the saddle-point approximation (SPA), as in the coordinate-space approach~\cite{cai2025prl}.
First, we apply the SPA to the functional integral over $\xi$.
The variation of the momentum-space action $S_p[x,p;t]=\int_0^t[-x\dot p-H(x,p)]\mathrm{d}t'$ with respect to the phase-space path $\xi(t')=\{x(t'),p(t')\}$ yields the complex canonical equations
$\dot{x}={\partial H}/{\partial p},
\dot{p}=-{\partial H}/{\partial x}$.
Second, applying the SPA to the integral over $p_i$ imposes
$x(t)=x(0)$, implying that the classical orbits are periodic in the complex phase space.
Finally, the saddle point of the time integral is determined by ${\partial}\left[S_p(t)+Et\right]/{\partial t}=0.$
This condition fixes the orbit to the energy shell,
$H(x(t'),p(t'))=E$,
and the saddle-point value of $t$ is the period $T(E)$, which is generally complex.
As shown in Appendix~\ref{App.A3}, after three steps of SPAs, the trace formula can be expressed in terms of contour integrals along all periodic orbits $O$ as
\begin{equation}\label{Gp_semi}
G(E)=iT(E)\frac{\exp\left[i\left(-\oint_{O}x\,\mathrm{d}p-2\pi\mu\right)\right]}
{1-\exp\left[i\left(-\oint_{O}x\,\mathrm{d}p-2\pi\mu\right)\right]},
\end{equation}
where $T(E)$ is the periodicity of the
orbit for a given energy $E$ and
$-4\mu$ defines the Maslov index.
Equation~\eqref{Gp_semi} is the dual form of the Gutzwiller trace formula in coordinate space.

The poles of Eq.~\eqref{Gp_semi} yield the momentum-space quantization condition
\begin{equation}\label{QCp}
-\oint_{O}x\,\mathrm{d}p
=
2\pi(n+\mu),
\qquad n\in\mathbb Z .
\end{equation}
The corresponding coordinate-space quantization condition is~\cite{cai2025prl}
\begin{equation}\label{QCx}
  \oint_{O}p\,\mathrm{d}x
=
2\pi(n+\mu),
\qquad n\in\mathbb Z .
\end{equation}
We now compare the momentum- and coordinate-space quantization conditions.
For contractible periodic orbits satisfying $p(0)=p(T), x(0)=x(T)$, the two quantization conditions are equivalent because
\begin{equation}\label{px}
\left.\oint p\dd x=\oint -x\dd p+px\right|_{p(0),x(0)}^{p(T),x(T)}=\oint -x\dd p.
\end{equation}
For non-contractible periodic orbits, however, either $p(T)\neq p(0)$ or $x(T)\neq x(0)$, so the boundary term no longer vanishes. 
As an example, consider a Brillouin-zone-winding orbit with $p(T)=p(0)+2\pi, x(T)=x(0)$. The two contour integrals are then related by
\begin{equation}\label{13}
    \left.\oint p\dd x=\oint -x\dd p+px\right|_{p(0),x(0)}^{p(T),x(T)}=\oint -x\dd p+2\pi x(0).
\end{equation}
The contractible and noncontractible periodic orbits mentioned above can be distinguished by their phase-space winding numbers \cite{contractibility}.

 Equation~\eqref{13} shows that the two contour integrals differ by a boundary term that explicitly depends on the starting point along the winding orbit. 
{ This apparent discrepancy
reflects the global topology of momentum space. For a Brillouin-zone-winding orbit, the contour integral \(-\oint x\,\mathrm{d}p\) is globally well defined. 
By contrast, since \(p\) cannot be chosen as a globally single-valued variable on the momentum cylinder, the bare contour integral \(\oint p\,\mathrm{d}x\) is not globally well defined.
When the orbit crosses the chosen Brillouin-zone boundary, a boundary term  $-2\pi x(0)$ arises from the global topology of momentum space.
Once this term is included, the coordinate- and momentum-space formulations
yield the same quantization condition, naturally expressed in the
momentum-space form of Eq.~\eqref{QCp}.
A detailed derivation of this result is provided in Appendix~\ref{globalQC}.
 An analogous argument applies to real-space-winding orbits, with the roles of coordinate and momentum interchanged. 
In this case, the coordinate-space quantization condition Eq.~\eqref{QCx} provides the globally well-defined rule, consistent with the examples discussed in Ref.~\cite{cai2025prl}.
}

Having identified the appropriate quantization condition for each type of periodic orbit, we next establish the quantum-classical correspondence in momentum-space formulation, relating the real or complex character of individual energy levels to the symmetry properties of the corresponding orbits.
As shown in Appendix~\ref{App.A4},
if a periodic orbit $O_1(t)$ with energy $E$ and period $T$ is a solution of Hamilton's canonical equations,  $O_2(t)=\xi_\eta(O_1^*(\pm t))$ is also a solution of the equations, with energy $E^*$ and period $T^*$. Moreover, the contour integrals of two orbits satisfy:
\begin{equation}
      \oint_{O_2(t^*)} -x_2(t^*) \mathrm{d}p_2(t^*)=\left(\oint_{O_1(t)} -x_1(t) \mathrm{d}p_1(t)\right)^*.
\end{equation}

\begin{figure}
    \centering
    \includegraphics[width=0.6\linewidth]{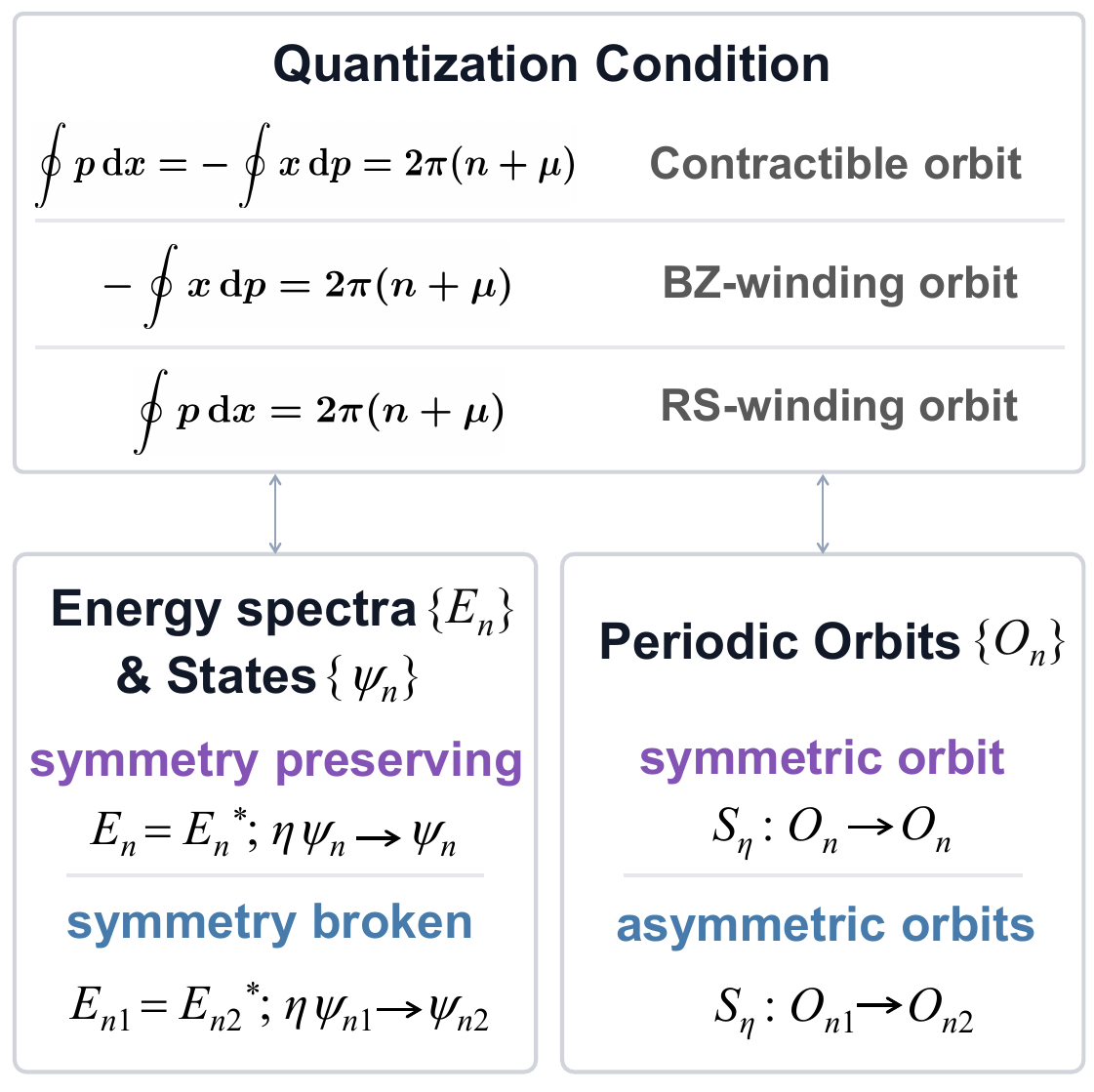}
    \caption{A more complete quantum-classical correspondence for non-Hermitian symmetry breaking. Energy spectra and quantum states are related to semiclassical periodic orbits through quantization conditions. Contractible periodic orbits can be quantized equivalently in coordinate or momentum space, whereas Brillouin-zone-winding orbits (BZ-winding orbits) and real-space-winding orbits (RS-winding orbits) require the momentum-space and coordinate-space quantization conditions, respectively. The symmetry properties of the orbits further determine the spectral character: $S_\eta$-symmetric orbits correspond to symmetry-preserving states with real eigenenergies, while asymmetric orbits forming $S_\eta$-symmetric pairs correspond to symmetry-broken states with complex-conjugate eigenenergies.
    }
    \label{fig1}
\end{figure}

Together with the quantization condition, these symmetry relations provide a
classification of nondegenerate levels according to the symmetry of the semiclassical orbit:  
\begin{enumerate}
\item  When two orbits $O_1$ and $O_2$ describe the same orbit, i.e., the orbit is $S_\eta$ symmetric, the quantized energy is real. 
\item When $O_1$ and $O_2$ are distinct but related by $S_\eta$, they form an $S_\eta$-related orbit pair, and their quantized energies occur as complex-conjugate pairs, $E_{n2}=E_{n1}^*$.
\end{enumerate}
This establishes the quantum-classical correspondence in the momentum-space formulation, consistent with its coordinate-space counterpart. Combined with the coordinate-space formulation, it completes the quantum-classical correspondence framework for non-Hermitian symmetry and spectral properties, as summarized in Fig.~\ref{fig1}.
\section{Boundary-induced real-to-complex transition of Bloch oscillations}
In Ref.~\cite{cai2025prl}, we applied the coordinate-space quantization condition to several examples, thereby demonstrating its validity. Here we consider the boundary-induced real-to-complex transition of Bloch oscillations. 
Because a particle undergoing Bloch oscillations winds around the Brillouin zone in momentum space, {the conventional coordinate-space formulation requires an additional boundary term, whereas the momentum-space formulation is globally well defined in its natural form. } 
It therefore provides an ideal test of the momentum-space quantization condition.
We begin with Bloch oscillations in a one-dimensional lattice (with the lattice constant set to unity), whose band dispersion $h(p)$ is periodic in the Brillouin zone. The Hamiltonian reads
\begin{equation}
    \mathcal{H}=h(\hat{p})+F\hat{x},
\end{equation}
where $F$ is a uniform external force.
We first focus on the bulk Bloch motion and neglect boundary effects; boundary reflections in finite lattices will be incorporated later in a concrete example.
The corresponding classical Hamiltonian is
$H(x,p)=h(p)+Fx$, from which Hamilton's equations follow:
\begin{equation}\label{HJ_eq}
    \dot{x}=\frac{\partial H}{\partial p}
    =\frac{\partial h(p)}{\partial p},
    \qquad
    \dot{p}=-\frac{\partial H}{\partial x}=-F.
\end{equation}
For the initial conditions $p(0)=p_0$ and $x(0)=[E-h(p_0)]/F$, the solution of Eq.~\eqref{HJ_eq} yields
\begin{equation}
    p(t)=p_0-Ft,
    \qquad
    x(t)=\frac{E-h\!\left(p(t)\right)}{F},
    \label{general_motion}
\end{equation}
where $x(0)$ is expressed in terms of $p_0$ and the energy $E$ through the relation $E=h(p_0)+Fx_0$.
Equation~\eqref{general_motion} shows that the momentum evolves uniformly and
winds across the Brillouin zone during the Bloch period
$T={2\pi}/{F}$.
Since the band dispersion is periodic, $h(p+2\pi)=h(p)$, the coordinate also
satisfies $x(t+T)=x(t)$, exhibiting Bloch oscillation in complex coordinate space. 

Having obtained the orbits of Bloch oscillations, we can determine the quantized energies by applying the momentum-space quantization condition Eq.~\eqref{QCp}.
To elucidate the relation between the quantized energy levels and the band dispersion,
we expand $h(p)$ as a Fourier series,
\begin{equation}\label{hp}
    h(p)=h_0+\sum_{\ell=1}^{\infty}
    \left(a_\ell\cos \ell p+b_\ell\sin \ell p\right),
\end{equation}
where the Fourier coefficients are generally complex in a non-Hermitian system. 
Substituting the orbit in Eq.~\eqref{general_motion} and the Fourier expansion
in Eq.~\eqref{hp} into the contour integral in Eq.~\eqref{QCp}, we obtain
\begin{equation}\label{21}
-\oint x\mathrm{d}p=\frac{1}{F}\int_{p_0-2\pi}^{p_0}\left[E-h(p)\right]\mathrm{d}p=\frac{2\pi}{F}\left(E-h_0\right),
\end{equation}
where we have used the fact that all
nonzero Fourier harmonics in Eq.~\eqref{hp} vanish upon integration over the Brillouin zone. 
The resulting quantized energies are
\begin{equation}
    E_n=h_0+Fn,
    \label{general_En}
\end{equation}
where the Maslov index is taken to be $\mu=0$, since the orbit encounters neither turning points nor reflections in momentum space.
The results show that the detailed form of the band dispersion enters the quantized spectrum only through the constant offset $h_0$.

\begin{figure}
    \centering
    \includegraphics[width=0.8\linewidth]{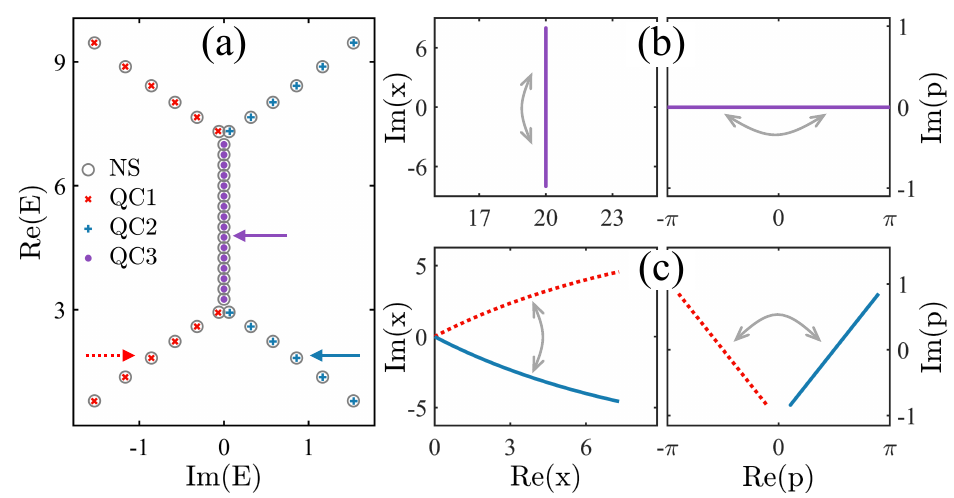}
    \caption{
Boundary-induced real-to-complex spectral transition of Bloch oscillations in
a finite non-Hermitian lattice.
(a) Comparison between the numerical spectrum (NS), obtained by {solving the matrix eigenvalue problem for the finite chain Hamiltonian under open boundary conditions,}
 and the semiclassical energy levels
obtained from the quantization conditions QC1--QC3 applied to the periodic orbits shown in (b) and (c).
{The detailed calculation of the quantization conditions for orbits $O_1$ and $O_2$ is presented in Appendix~B.}
(b) The $S_{\mathsf T}$-symmetric Brillouin-zone-winding orbit in complex
phase space corresponding to QC3.
(c) A representative pair of $S_{\mathsf T}$-related orbits reflected
from the left boundary, corresponding to QC1 and QC2. The gray double-headed
arrows indicate the $S_{\mathsf T}$ symmetry operation.
The parameters are $t_0=1$, $L=41$, and $F=0.25$, with the two boundaries located at $x=0$ and $x=L$, corresponding to  lattice sites $N=L-1=40$.
}
    \label{fig2}
\end{figure}

Having established the semiclassical description of Bloch oscillations for a general dispersion, we turn to a concrete model:
\begin{equation}\label{Hbloch}
    \mathcal{H}(\hat{x},\hat{p})
    =2it_0\sin\hat{p}+F\hat{x},
\end{equation}
where $t_0$ characterizes the magnitude of the imaginary dispersion.
For clarity, we take $t_0>0$ and $F>0$ and consider a finite lattice bounded by $x=0$ and $x=L$. 
We further assume that $L$ is sufficiently large to permit the existence of bulk semiclassical orbits that can complete a full Bloch oscillation before reaching either boundary.
For complex-valued momentum, we restrict our discussion to the first Brillouin zone, defined by $-\pi\leq\operatorname{Re}p\leq\pi$. 
The Hamiltonian obeys the time-reversal $\mathcal{T}$ constraint
\begin{equation}
    \mathcal{H}^{*}(\hat{x},-\hat{p})=\mathcal{H}(\hat{x},\hat{p}),
\end{equation}
or $\mathsf{T}$-PHS in the standard pseudo-Hermitian notation~\eqref{hs}, where $\mathsf{T}$ denotes matrix transposition.
As shown in Fig.~\ref{fig2}(a), the eigenenergies of this Hamiltonian are complex in the upper and lower parts of the spectrum, whereas they remain real in the central spectral region.
This spectral structure can be understood within the semiclassical framework.
The corresponding classical symmetry $S_{\mathsf{T}}$ is expressed as
\begin{equation}
    H(x,-p)=H^{*}(x^{*},p^{*}),
\end{equation}
which yields the $S_{\mathsf{T}}$-symmetric pairs of orbits denoted as
$O_1(t)=\{x(t),p(t)\}$ and $O_2(t)=\left\{x^{*}(-t),-p^{*}(-t)\right\}.$
Physically, the semiclassical orbits fall into two classes: bulk orbits that complete a full Bloch oscillation, and boundary-reflected orbits for which the Bloch oscillation is interrupted by the boundary.

For bulk orbits $O$, the semiclassical solution is
\begin{equation}\label{motion}
    x(t)=\frac{1}{F}\left[E-2it_0\sin(Ft)\right],
    \qquad
    p(t)=\pi-Ft,
\end{equation}
where the initial momentum is taken as $p_0=\pi$.
For a real energy $E$, this orbit $O$ satisfies $S_{\mathsf{T}}$ symmetry, i.e.,
\begin{equation}
    x^{*}(-t)=x(t),
    \qquad
    -p^{*}(-t)=p(t)
    \quad (\mathrm{mod}\ 2\pi),
\end{equation}  as shown in Fig.~\ref{fig2}(b). Note that for real time $t$, the Bloch oscillation occurs along the imaginary coordinate axis, in contrast to the conventional Hermitian counterpart.
Applying the momentum-space quantization condition~\eqref{QCp} to this orbit yields real energy levels
$E_n=Fn$, as shown by the purple points in Fig.~\ref{fig2}(a). The eigenenergies obtained from the quantization condition agree well with the numerical diagonalization results, confirming the validity of our theory.
The physical interpretation of these energy levels is that, for a particle undergoing a Brillouin-zone winding, the gain and loss effects induced by the imaginary dispersion \(h(p)=2it_0\sin p\) compensate each other over one period, thereby ensuring that the corresponding eigenvalue remains real.
This mechanism is analogous to
that in the \(\mathcal{PT}\)-symmetric double-well model \cite{cai2025prl}, where high-energy orbits explore both gain and loss regions in coordinate space so that the two effects compensate over one period.

However, once the particle is reflected by the boundary during the Bloch oscillation, it can no longer complete a full winding around the Brillouin zone. Instead, the motion is confined to only part of the Brillouin zone, so that the gain and loss accumulated along the trajectory are no longer balanced. Consequently, the boundary-reflected orbit \(O_1\), subject to the loss, breaks the \(S_{\mathsf T}\) symmetry, while its \(S_{\mathsf T}\)-symmetric partner \(O_2\) experiences the corresponding gain. They are represented by the red dashed and blue solid trajectories, respectively, in Fig.~\ref{fig2}(c).

We first identify the boundary-reflected orbits and then apply the quantization condition to them. Since the force $F$ creates a linear potential gradient across the system, the boundary-reflected orbits localized near the two boundaries naturally correspond to the two ends of the energy spectrum, as confirmed in Fig.~\ref{fig2}(a). On the original real-time contour, however, Eq.~\eqref{motion} shows that the Bloch oscillation proceeds along the imaginary coordinate axis, obscuring the role of the real-space boundaries. While this description suffices for Bloch-oscillation orbits, it is inadequate for boundary-reflected ones. To capture the boundary effect, one should analytically continue the time contour into the complex plane, where the propagator naturally selects the dominant orbits.
This continuation allows orbits to reach and reflect from the boundary during the Bloch oscillation, giving rise to the two boundary-reflected orbits \(O_1\) and \(O_2\).
{ It should be noted that the boundary-reflected orbits and the Bloch orbits are complex saddle-point trajectories, which are distinct from the ordinary real-time and -space trajectories in Hermitian systems.}
 From the path-integral perspective, both complete Bloch-oscillation and boundary-reflected orbits contribute to the propagator.
The key is whether boundary-reflected orbits remain an exponentially small correction to the bulk Bloch oscillations, or instead become the dominant contribution that dictates the spectrum.
As shown in Appendix~\ref{App.B}, this distinction can be identified by examining the contributions of the Bloch-oscillation orbit $O$ and boundary-reflected orbit $O_2$ to the propagator. In the momentum-space path integral, an orbit with energy $E$ contributes a factor $\exp(-i\int x\,\mathrm{d}p)$, whose magnitude is given by $\left|\exp\left(-i\int x\,\mathrm{d}p\right)\right|=\exp\left(\operatorname{Im}\int x\,\mathrm{d}p\right)$.
Therefore, a negative \(\operatorname{Im}\int x\,\dd p\) suppresses the orbit contribution, whereas a positive value enhances it.
Here, for Bloch-oscillation orbits winding across the whole Brillouin zone, we have $\operatorname{Im}\oint_{O}x\,\dd p=0$. Consequently, the boundary-reflected orbit remains exponentially suppressed when
$\operatorname{Im}\oint_{O_{2}}x\,\dd p<0$, and the complete Bloch oscillation remains the dominant trajectory. Conversely, when $\operatorname{Im}\oint_{O_{2}}x\,\dd p>0$,
the contribution of the boundary-reflected orbit is enhanced and becomes the dominant one governing the semiclassical quantization.

For concreteness, we focus on the left boundary in the following discussion.
By evaluating the semiclassical weight of \(O_2\), we find that the
boundary-reflected contribution is suppressed for large positive \(E\), where the center of the Bloch-oscillation orbit $n_c=E/F$ lies far from the left boundary.
As \(E\) decreases, the orbit
becomes closer to the left boundary and correspondingly, the boundary-reflected contribution gradually increases. 
When the energy decreases below the critical value \(E^l_c\), determined by
\begin{equation}\label{critical}
-\operatorname{Im}\oint_{O_{2}}x(E^l_c,p)\,\mathrm{d}p=\operatorname{Im}\oint_{O_{2}}p(E^l_c,x)\,\mathrm{d}x=0,
\end{equation}
the boundary-reflected orbits become the dominant semiclassical contributions
governing the quantized energy levels.
{Mathematically, this transition condition corresponds to an anti-Stokes crossover.}
The solution of Eq.~\eqref{critical} yields
\begin{equation}
    E_c^{l}\approx 3.018\,t_0.
\end{equation}
In this energy regime, the eigenenergies
in the lower and upper halves of the complex-energy plane are governed
by the orbits $O_1$ and $O_2$ respectively, shown in Appendix~\ref{App.B}:
\begin{equation}
\begin{cases}
\displaystyle
\oint_{O_1}p\,\mathrm{d}x
=
2\pi\left(n+\dfrac{3}{4}\right),
& \operatorname{Im}E<0,
\\[3mm]
\displaystyle
\oint_{O_2}p\,\mathrm{d}x
=
2\pi\left(n+\dfrac{3}{4}\right),
& \operatorname{Im}E>0.
\end{cases}
\end{equation}
Here, the phase correction $2\pi\times3/4=3\pi/2$ in the quantization
condition consists of a phase shift $\pi$ from reflection at the hard
boundary and a phase shift $\pi/2$ from the smooth turning point
generated by the linear potential.
{A detailed discussion about Maslov phase can be found in Ref.~\cite{brack2003semiclassical}.}
 Since all boundary-reflected orbits are contractible, the equivalence relation in Eq.~\eqref{px} has been used.
The eigenvalues associated with the right-boundary-reflected orbits can be obtained by an analogous analysis.
As shown in Fig.~\ref{fig2}(a), the spectra obtained from the quantization conditions agree well with the numerical solutions of the eigenvalue equation, supporting the validity of the present semiclassical description.

\section{Conclusion}
In summary, we establish a quantum–classical correspondence for non-Hermitian symmetry breaking based on the momentum-space path-integral formulation. We derive the corresponding trace formula and quantization conditions, providing an essential complement to the real-space path-integral approach and extending the applicability of the theoretical framework to a broader class of non-Hermitian systems. 
{Furthermore, we demonstrate that orbit topology determines the natural formulation of semiclassical quantization. For contractible
periodic orbits in phase space, the coordinate- and momentum-space quantization conditions are directly equivalent. For noncontractible winding orbits, the winding direction determines which condition is
globally well defined in its conventional form: real-space and Brillouin-zone winding naturally select coordinate- and momentum-space quantization, respectively. The dual formulation requires an additional boundary term to recover the same quantization condition.}
Finally, we demonstrate the validity of our theory by studying the boundary-induced real-to-complex transition of Bloch oscillations in a finite non-Hermitian lattice.





\vspace{6pt} 


\funding{This work was supported by the National Natural Science Foundation of China (No. 12574051), the Natural Science Foundation of Jiangsu Province (Nos. BK20250008, BK20233001, BK20253009), the Fundamental Research Funds for the Central Universities (No. 2026010801), and the National Key Projects for Research and Development of China (No. 2022YFA1204701).}

\dataavailability{The data that support the findings of this study are available from the corresponding author upon reasonable request.}

\conflictsofinterest{The authors declare no conflicts of interest.} 


\abbreviations{Abbreviations}{
The following abbreviations are used in this manuscript:\\

\noindent 
\begin{tabular}{@{}ll}
$\eta$-PHS & $\eta$-pseudo-Hermitian symmetry \\
SPA & saddle point approximation
\end{tabular}
}

\appendixtitles{yes} 
\appendixstart
\appendix

\setcounter{equation}{0}
\renewcommand{\theequation}{A\arabic{equation}}
\renewcommand{\theHequation}{appendix.A.\arabic{equation}}

\setcounter{equation}{0}
\setcounter{figure}{0}
\setcounter{table}{0}

\renewcommand{\theequation}{A\arabic{equation}}
\renewcommand{\thefigure}{A\arabic{figure}}
\renewcommand{\thetable}{A\arabic{table}}
\section[\appendixname~\thesection]{Quantum-classical correspondence principle through momentum path integral}\label{App.A}
In this part, we provide the technical derivations and additional analysis supporting the momentum-space quantum-classical correspondence developed in the main text.
The organization is as follows. 
In Appendix A.1, we derive the constraint imposed by \(\eta\)-pseudo-Hermiticity on the momentum-space propagator. 
In Appendix A.2, we construct the momentum-space complex path integral and obtain the corresponding
classical \(S_{\eta}\) symmetry. 
In Appendix A.3, we derive the momentum-space non-Hermitian Gutzwiller trace formula and the associated quantization condition, and clarify its relation to the coordinate-space version for winding
orbits. 
{In Appendix~A.4, we discuss the quantization of Brillouin-zone-winding orbits in terms of momentum-space topology and the global properties of the canonical one-forms.}
In Appendix A.5, we show how the symmetry of semiclassical orbits constrains the real-or-complex nature of the quantized energy levels. 

\subsection[\appendixname~\thesubsection]{Restriction on the momentum-space propagator by \texorpdfstring{$\eta$}{eta}-pseudo-Hermiticity}\label{App.A1}
The $\eta$-pseudo-Hermitian symmetry ($\eta$-PHS) of the Hamiltonian is defined by
\begin{equation}\label{A1}
    \eta \mathcal{H}\eta^{-1}=\mathcal{H}^{\dagger} .
\end{equation}
In the momentum representation, we introduce the propagator
\begin{equation}\label{gpdef}
    G(p_f,p_i,t)=\langle p_f|e^{-i\mathcal{H}(\hat{x},\hat{p})t}|p_i\rangle,
\end{equation}
where $|p_i\rangle$ and $|p_f\rangle$ are biorthonormal momentum eigenstates. The propagator associated with the $\eta$-transformed Hamiltonian is
\begin{equation}\label{getapdef}
    G_\eta(p_f,p_i,t)=\langle p_f|e^{-i\mathcal{H}(\hat{x}_\eta,\hat{p}_\eta)t}|p_i\rangle.
\end{equation}
The coordinate and momentum operators under the similarity transformation
$\eta$ are denoted as
\begin{equation}\label{etaop}
    \hat{x}_\eta=\eta\hat{x}\eta^{-1},\qquad
    \hat{p}_\eta=\eta\hat{p}\eta^{-1}.
\end{equation}
Equation~\eqref{A1} imposes the momentum-space restriction
\begin{equation}\label{gprestrict}
    G_{\eta}(p_f,p_i,t)=G^*(p_i^*,p_f^*,-t^*) .
\end{equation}
This relation follows directly from
\begin{equation}
\begin{aligned}
G_{\eta}(p_f,p_i,t)
&=\langle p_f|e^{-i\mathcal{H}(\hat{x}_\eta,\hat{p}_\eta)t}|p_i\rangle
 =\langle p_f|e^{-i\mathcal{H}^{\dagger}(\hat{x},\hat{p})t}|p_i\rangle \\
&=\left[\langle p_i|\left(e^{-i\mathcal{H}^{\dagger}(\hat{x},\hat{p})t}\right)^{\dagger}|p_f\rangle\right]^*
 =\left[\langle p_i|e^{i\mathcal{H}(\hat{x}^{\dagger},\hat{p}^{\dagger})t^*}|p_f\rangle\right]^* \\
&=G^*(p_i^*,p_f^*,-t^*) .
\end{aligned}
\end{equation}
Here we have used the momentum-space biorthonormal relations
\begin{equation}
\hat{p}|p_i\rangle=p_i|p_i\rangle,\qquad
\langle p_i|\hat{p}^{\dagger}=\langle p_i|p_i^*,\qquad
\langle p_f|\hat{p}=\langle p_f|p_f,\qquad
\hat{p}^{\dagger}|p_f\rangle=p_f^*|p_f\rangle .
\end{equation}
It should be noted that the Hermitian conjugation in Eq.~\eqref{A1} acts on the Hamiltonian as an operator and does not act on the individual variables $\hat{x}$ and $\hat{p}$.

The same restriction can also be understood by analytic continuation. For real $p$, if two analytic functions $f(p)$ and $g(p)$ satisfy $f(p)=g^*(p)$, then their Taylor coefficients obey $f^{(n)}(0)=[g^{(n)}(0)]^*$, and hence
\begin{equation}
    f(p)=\sum_n\frac{f^{(n)}(0)}{n!}p^n
    =\sum_n\left[\frac{g^{(n)}(0)}{n!}(p^*)^n\right]^*
    =g^*(p^*) .
\end{equation}
Applying this property to the momentum-space propagator, Eq.~\eqref{A1} gives, for real momenta and real time,
\begin{equation}
G_{\eta}(p_f,p_i,t)
=\langle p_f|e^{-i\mathcal{H}(\hat{x}_\eta,\hat{p}_\eta)t}|p_i\rangle
=\left[\langle p_i|e^{i\mathcal{H}(\hat{x},\hat{p})t}|p_f\rangle\right]^*
=G^*(p_i,p_f,-t).
\end{equation}
\subsection{Momentum-space complex path integral and classical \texorpdfstring{$S_\eta$}{S eta} symmetry}\label{App.A2}
{
We start from the momentum-space propagator
\begin{equation}
G(p_f,p_i,t)=\langle p_f|e^{-i\mathcal{H}t}|p_i\rangle ,
\end{equation}
where \(p_i\), \(p_f\), and \(t\) have been analytically continued to complex values.
Although these external variables are complex, at each intermediate time slice we insert the conventional completeness relations defined on the real axes,
\begin{equation}
\int_{\mathbb R}\mathrm{d}p\,|p\rangle\langle p|= I,
\qquad
\int_{\mathbb R}\mathrm{d}x\,|x\rangle\langle x|= I.
\end{equation}
The complex endpoint momenta enter as external parameters through
\(p_0=p_i\) and \(p_N=p_f\). 
To prepare for the subsequent discussion in complexified phase space, we deform the original integration contours of the discretized path integral, \(\int_{\mathbb R}\mathrm{d}p_n\) and
\(\int_{\mathbb R}\mathrm{d}x_n\), into contours
\(\int_{\Gamma_{p,n}}\mathrm{d}p_n\) and
\(\int_{\Gamma_{x,n}}\mathrm{d}x_n\) in the complex plane.
There is freedom in choosing the deformed contours $\Gamma_{p,n}$ and $\Gamma_{x,n}$, provided that the integrand remains analytic, no singularities are crossed, and the endpoint and convergence conditions are preserved.
We thereby obtain a discretized path-integral representation over complex phase-space trajectories
\begin{equation}
\begin{aligned}
G(p_f,p_i,t)
=
\int
\prod_{n=1}^{N-1}\mathrm{d}p_n
\prod_{n=1}^{N}\frac{\mathrm{d}x_n}{2\pi}
\exp\left\{
i\sum_{n=1}^{N}
\left[
-x_n(p_n-p_{n-1})
-\epsilon H(x_n,p_n)
\right]
\right\}.
\end{aligned}
\end{equation}
Taking the continuum limit, we obtain the path-integral representation of $G(p_f,p_i,t)$ in complex phase space:
\begin{equation}
\begin{aligned}
G(p_f,p_i,t)
=\int_{p_i}^{p_f}\mathcal{D}[\xi]
\exp\left\{i\int_0^t[-x\dot p-H(x,p)]\mathrm{d}t'\right\},
\end{aligned}
\end{equation}
where $\xi(t)=\{x(t),p(t)\}^{\mathsf T}$, $N\epsilon=t$, and $\mathcal{D}[\xi]=\prod_{n=1}^{N-1}\mathrm{d}p_n
\prod_{n=1}^{N}\mathrm{d}x_n/(2\pi)$.
}
Compared with the coordinate-space action, the symplectic term has the dual form $-x\dot p$, which is the natural form when the endpoints are fixed in momentum space.
For the $\eta$-transformed propagator, the same construction gives
\begin{equation}\label{pathpeta}
\begin{aligned}
G_{\eta}(p_f,p_i,t)
&=\int\prod_{n=1}^{N-1}\dd p_n\prod_{n=1}^{N} \frac{\dd x_n}{2\pi}
\exp\left\{i\sum_{n=1}^{N}\left[-x_n(p_n-p_{n-1})
-\epsilon H(x_\eta(x_n,p_n),p_\eta(x_n,p_n))\right]\right\} \\
&=\int_{p_i}^{p_f}\mathcal{D}[\xi]
\exp\left\{i\int_0^t[-x\dot p-H(x_\eta(x,p),p_\eta(x,p))]\dd t'\right\} .
\end{aligned}
\end{equation}
Here $x_\eta(x,p)$ and $p_\eta(x,p)$ are the classical symbols induced by the operator transformation in Eq.~\eqref{etaop}.

We now translate the propagator constraint~\eqref{gprestrict} into the path-integral language. It gives
\begin{equation}\label{pathsym0}
\begin{aligned}
&\int_{p_i}^{p_f}\mathcal{D}[\xi]
\exp\left\{i\int_0^t[-x\dot p-H(x_\eta(x,p),p_\eta(x,p))]\dd t'\right\} \\
&\qquad =\left(\int_{p_f^*}^{p_i^*}\mathcal{D}[\xi]
\exp\left\{i\int_0^{-t^*}[-x\dot p-H(x,p)]\dd t'\right\}\right)^* .
\end{aligned}
\end{equation}
The right-hand side can be transformed by changing variables $x\rightarrow x^*$, $p\rightarrow p^*$, and $t'\rightarrow t'^*$:
\begin{equation}
\begin{aligned}
&\left(\int_{p_f^*}^{p_i^*}\mathcal{D}[\xi]
\exp\left\{i\int_0^{-t^*}[-x\dot p-H(x,p)]\dd t'\right\}\right)^* \\
&\qquad =\int_{p_f}^{p_i}\mathcal{D}[x]\mathcal{D}[p]
\exp\left\{-i\int_0^{-t}\left[-x\frac{\dd p}{\dd t'}-H^*(x^*,p^*)\right]\dd t'\right\}.
\end{aligned}
\end{equation}
Interchanging the integration limits and shifting the time contour then gives
\begin{equation}
\int_{p_i}^{p_f}\mathcal{D}[\xi]
\exp\left\{i\int_0^t[-x\dot p-H^*(x^*,p^*)]\dd t'\right\}.
\end{equation}
Therefore Eq.~\eqref{pathsym0} becomes
\begin{equation}\label{pathsym}
\begin{aligned}
&\int_{p_i}^{p_f}\mathcal{D}[\xi]
\exp\left\{i\int_0^t[-x\dot p-H(x_\eta(x,p),p_\eta(x,p))]\dd t'\right\} \\
&\qquad =\int_{p_i}^{p_f}\mathcal{D}[\xi]
\exp\left\{i\int_0^t[-x\dot p-H^*(x^*,p^*)]\dd t'\right\}.
\end{aligned}
\end{equation}
{In contrast to Eq.~\eqref{A1}}, the complex conjugation in Eq.~\eqref{seta} acts on the c-number variables $x^*$ and $p^*$.
{
    Since Eq.~\eqref{pathsym} is valid for arbitrary $t$ and arbitrary initial and final momenta $p_i$ and $p_f$, the corresponding semiclassical Hamiltonian obeys the following $S_\eta$ symmetry:
\begin{equation}\label{seta}
    S_\eta: H(x_\eta(x,p),p_\eta(x,p))=H^*(x^*,p^*) .
\end{equation}
The proof is as follows:
Considering the single-time-slice short-time
limit \(t=\epsilon\rightarrow0\), Eq.~\eqref{pathsym} becomes
\begin{equation}\label{A18}
\begin{aligned}
        \int  e^{-i(p_f-p_i)x-i\epsilon H(x_\eta(x,p_f),p_\eta(x,p_f))}dx&=\int  e^{-i(p_f-p_i)x-i\epsilon H^*(x^*,p_f^*)}dx\\
         \int  e^{-i(p_f-p_i)x}\left[1-i\epsilon H(x_\eta(x,p_f),p_\eta(x,p_f))\right]dx&=\int  e^{-i(p_f-p_i)x}\left[1-i\epsilon H^*(x^*,p_f^*)\right]dx\\
         \int  e^{-i(p_f-p_i)x}\left[H(x_\eta(x,p_f),p_\eta(x,p_f))-H^*(x^*,p_f^*)\right]dx&=0.    
\end{aligned}
\end{equation}
Since Eq.~\eqref{A18} holds for arbitrary $p_i$ and $p_f$, the momentum difference can be chosen arbitrarily. This implies that all Fourier components of the quantity in square brackets vanish, and hence the quantity in square brackets must vanish identically.
This proves Eq.~\eqref{seta}.}

{A direct derivation of the classical Hamiltonian symmetry from the
operator \(\eta\)-pseudo-Hermiticity relation proceeds as follows.
We first assume that the coordinate and momentum are real and derive the symmetry relation of the semiclassical Hamiltonian on the real
phase space. We then analytically continue this relation into complex phase space.
The semiclassical Hamiltonian can be obtained from the following equation:
\begin{equation}
H(x,p)=\lim_{\hbar\rightarrow 0}\frac{\langle p| \mathcal{H}(\hat{x}, \hat{p})|x\rangle}{\langle p \mid x\rangle}=\lim_{\hbar\rightarrow 0}\frac{\langle x| \mathcal{H}(\hat{x}, \hat{p})|p\rangle}{\langle x \mid p\rangle}.
\end{equation}
From $\eta$-PHS of the Hamiltonian, we have
\begin{equation}
    \frac{\langle p| \mathcal{H}(\hat{x}_{\eta}, \hat{p}_{\eta})|x\rangle}{\langle p \mid x\rangle}=  \frac{\langle p| \mathcal{H}^{\dagger}(\hat{x}, \hat{p})|x\rangle}{\langle p \mid x\rangle}.
\end{equation}
Making use of the identity
\[
\frac{
\langle p|\hat{\mathcal H}^{\dagger}|x\rangle
}{
\langle p|x\rangle
}
=
\left[
\frac{
\langle x|\hat{\mathcal H}|p\rangle
}{
\langle x|p\rangle
}
\right]^*,
\]
and taking the principal semiclassical limit \(\hbar\to 0\), we obtain
\begin{equation}
    H(x_{\eta},p_{\eta})=H^*(x,p).
\end{equation}
Analytically continuing this relation to complex phase space yields the $S_\eta$ symmetry:
\begin{equation}
    H(x_{\eta},p_{\eta})=H^*(x^*,p^*),
\end{equation}
 which is Eq.~\eqref{seta}.
}

\subsection{Non-Hermitian trace formula in momentum space}\label{App.A3}
In this part, we generalize the Gutzwiller trace formula to momentum space for non-Hermitian systems. The derivation is the momentum-space dual of the coordinate-space construction and contains three saddle-point approximations applied to
\begin{equation}\label{GEp}
\begin{aligned}
G(E)
&=i\int \dd t\,e^{iEt}\int \dd p_i\,G_p(p_f=p_i,p_i,t) \\
&=i\int \dd t\,e^{iEt}\int \dd p_i\int_{p_i}^{p_i}\mathcal{D}[\xi]
\exp\left\{i\int_0^t[-x\dot p-H(x,p)]\dd t'\right\}.
\end{aligned}
\end{equation}
If momentum space is the Brillouin zone, this equality is understood modulo the Brillouin-zone period $2\pi$, i.e., $p(t)=p(0)+2\pi w$ with $w\in\mathbb Z$.

The first saddle-point approximation is applied to the path integral at fixed boundary momenta. We expand the momentum-space action
\begin{equation}\label{Spdef}
    S_p[x(t),p(t)]=\int_0^t[-x\dot p-H(x,p)]\dd t'
\end{equation}
around its saddle point. The first variation reads
\begin{equation}\label{deltaSp}
\delta S_p=\int_0^t\left[
\left(\dot{x}-\frac{\partial H}{\partial p}\right)\delta p
+\left(-\dot{p}-\frac{\partial H}{\partial x}\right)\delta x
\right]\dd t'-[x\delta p]_0^t .
\end{equation}
For fixed $p_i$ and $p_f$, the boundary term vanishes. The saddle-point condition $\delta S_p=0$ gives the complex Hamilton equations
\begin{equation}\label{eom}
    \dot{x}_{cl}=\frac{\partial H}{\partial p_{cl}},\qquad
    \dot{p}_{cl}=-\frac{\partial H}{\partial x_{cl}}.
\end{equation}
The subscript ``$cl$'' denotes the semiclassical solution and will be omitted when no confusion arises. The momentum-space propagator then reduces to the contribution from the classical orbit and its quantum fluctuations,
\begin{equation}\label{Gpfluc}
    G(p_f,p_i,t)\simeq e^{iS_p[x_{cl}(t'),p_{cl}(t')]}
    \Delta_p[x_{cl}(t'),p_{cl}(t')] .
\end{equation}
The fluctuation term is
\begin{equation}\label{Deltap}
\begin{aligned}
\Delta_p&=\int_{0}^{0}\mathcal{D}[x_d]\mathcal{D}[p_d]
\exp\left\{\frac{i}{2}\delta^2S_p[x_d(t'),p_d(t'),x_{cl}(t'),p_{cl}(t')]\right\},\\
\delta^2S_p&=\int_0^t\left[-2x_d\dot p_d-A_{cl}x_d^2-2B_{cl}x_dp_d-C_{cl}p_d^2\right]\dd t',\\
A_{cl}(t')&=\frac{\partial^2H}{\partial x_{cl}^2},\qquad
B_{cl}(t')=\frac{\partial^2H}{\partial x_{cl}\partial p_{cl}},\qquad
C_{cl}(t')=\frac{\partial^2H}{\partial p_{cl}^2}.
\end{aligned}
\end{equation}
Here $x_d=x-x_{cl}$ and $p_d=p-p_{cl}$ measure the deviations from the classical trajectory. The dual fluctuation calculation gives the compact semiclassical propagator
\begin{equation}\label{gpscl}
    G(p_f,p_i,t)=
    \sqrt{\frac{-1}{2\pi i\dot p_{cl}(t)\dot p_{cl}(0)\frac{\partial t}{\partial E_{cl}}}}
    e^{iS_{p,cl}(p_f,p_i,t)},
\end{equation}
where $S_{p,cl}(p_f,p_i,t)\equiv S_p[x_{cl},p_{cl}]$ and $E_{cl}=H(x_{cl},p_{cl})$ is the energy of the classical orbit. This is the analytic continuation of the corresponding Hermitian result to complex phase space, now written in the momentum representation.

We next impose the trace condition in momentum space. In Eq.~\eqref{GEp}, the paths are closed in the projected momentum coordinate, $p_f=p_i$. The second saddle-point approximation is applied to
\begin{equation}
    G(t)=\int \dd p_i\,G(p_i,p_i,t).
\end{equation}
The saddle point of the integrand is determined by
\begin{equation}\label{pitrace}
\partial_{p_i}S_{p,cl}(p_i,p_i,t)
=\partial_aS_{p,cl}(a,p_i,t)|_{a=p_i}
+\partial_bS_{p,cl}(p_i,b,t)|_{b=p_i}
=-x_{cl}(t)+x_{cl}(0)=0 .
\end{equation}
Thus the initial and final coordinates must also coincide, including both real and imaginary parts. The saddle points selected by the momentum trace are therefore periodic orbits in the full complex phase space.

The basic period of an orbit contributing to $G(t)$ need not be $t$ itself. It may be $t/n$ with $n=1,2,3,\ldots$. Taking repetitions into account, we obtain
\begin{equation}\label{Gpt}
\begin{aligned}
G(t)
&=\sum_\gamma\sum_{n=1}^{\infty}\oint\frac{\dd p_{cl}}{\dot p_{cl}}
\sqrt{\frac{-1}{2\pi i\frac{\dd t}{\dd E_{cl}}}}
\exp\left[i n S_{p,cl}^{\gamma}(t/n)\right] \\
&=\sum_\gamma\sum_{n=1}^{\infty}\frac{t}{n}
\sqrt{\frac{-1}{2\pi i\frac{\dd t}{\dd E_{cl}}}}
\exp\left\{i\left[n\oint_{O_{\gamma}}(-x_{cl}\dd p_{cl})-2\pi\mu_\gamma n-E_{cl}t\right]\right\}.
\end{aligned}
\end{equation}
Here $\gamma$ labels distinct families of periodic orbits, excluding different repetitions of the same primitive orbit. The contour $O_{\gamma}$ denotes the corresponding periodic orbit in phase space, and $S_{p,cl}^{\gamma}(t/n)$ is the action accumulated over one primitive period. The phase $-2\pi\mu_\gamma n$ arises from the poles of $\dd t/\dd E_{cl}$ associated with $\dot p_{cl}=0$, and $-4\mu_\gamma$ is the Maslov index.

Finally, we perform the third saddle-point approximation for
\begin{equation}
    G(E)=i\int \dd t\,e^{iEt}G(t).
\end{equation}
With the substitution $\tilde t=t/n$, Eq.~\eqref{Gpt} becomes
\begin{equation}\label{GpEbefore}
G(E)=i\sum_\gamma\sum_{n=1}^{\infty}\int \dd \tilde t\,
\sqrt{n}\tilde t
\sqrt{\frac{-1}{2\pi i\frac{\dd \tilde t}{\dd E_{cl}}}}
\exp\left\{in\left[E\tilde t+\oint_{O_{\gamma}}(-x_{cl}\dd p_{cl})-2\pi\mu_\gamma-E_{cl}\tilde t\right]\right\}.
\end{equation}
The saddle point is determined by
\begin{equation}
    E=-\left[\partial_{\tilde t}S_{p,cl}^{\gamma}(\tilde t)\right]_{\tilde t=T_\gamma}=E_{cl}(T_\gamma),
\end{equation}
which selects periodic orbits on the energy shell $E$. After the $\tilde t$ integration, the momentum-space trace formula becomes
\begin{equation}\label{tracep}
\begin{aligned}
G(E)
&=i\sum_\gamma\sum_{n=1}^{\infty}T_\gamma(E)
\exp[-i2\pi n\mu_\gamma]
\exp\left[in\oint_{O_{\gamma}}(-x_{cl}\dd p_{cl})\right] \\
&=i\sum_\gamma T_\gamma(E)
\frac{\exp\left\{i\left[\mathcal{J}_{\gamma}(E)-2\pi\mu_\gamma\right]\right\}}
{1-\exp\left\{i\left[\mathcal{J}_{\gamma}(E)-2\pi\mu_\gamma\right]\right\}},
\end{aligned}
\end{equation}
where $\mathcal{J}_{\gamma}(E)=-\oint_{O_{\gamma}}x\dd p$.
Equation~\eqref{tracep} constitutes the Gutzwiller trace formula in momentum space. 
When only a single family of periodic orbits is involved, the label $\gamma$ will be omitted for simplicity.

The poles of Eq.~\eqref{tracep} give the momentum-space quantization condition
\begin{equation}\label{quantp}
    -\oint_{O_{\gamma}}x\dd p=2\pi(n+\mu_\gamma),\qquad n\in\mathbb{Z}.
\end{equation}

{\subsection{Appropriate formulation of the quantization conditions for winding orbits}\label{globalQC}}
{
For Brillouin-zone-winding orbits, Eq.~\eqref{13} in the main text shows that the coordinate- and momentum-space contour integrals differ by a boundary term. 
To clarify the geometric origin of this term and
determine the appropriate quantization condition, we examine the global properties of the corresponding canonical one-forms in this part.
Owing to Brillouin-zone periodicity, the complex momentum space has the cylindrical topology $ S^{1}_{\operatorname{Re}p}\times\mathbb{R}_{\operatorname{Im}p}$.
Here, the real part of momentum is periodic, whereas its imaginary part is noncompact. 
The momentum \(p\) cannot be represented by a globally single-valued function on this cylinder since \(p\)
and \(p+2\pi\) describe the same physical momentum. 
Its differential \(\mathrm{d}p\), however, is globally well defined, since
\(\mathrm{d}(p+2\pi)=\mathrm{d}p\). 
Moreover, because the coordinate-space projection of the orbit is contractible, \(x\) can be chosen as a single-valued coordinate along the entire orbit.
Consequently, the canonical one-form
$    \vartheta_{p}=-x\,\mathrm{d}p$
is globally defined, whereas
$
    \vartheta_{x}=p\,\mathrm{d}x
$
is defined only locally.
Therefore, the quantization condition
\(\oint_O\vartheta_p=2\pi(n+\mu)\) remains globally well defined and directly applicable.}

{By contrast, formulating the quantization condition in terms of
\(\vartheta_x\) requires covering the momentum cylinder with two overlapping charts, which we take to be
$U_1:\ 0<\operatorname{Re}p_1<2\pi$ and
 $U_2:\ 2\pi-\delta<\operatorname{Re}p_2<2\pi+\delta,$
where \(\delta>0\) is taken to approach zero.
In these two charts, the local canonical one-forms are defined as $\vartheta_{1}=p_{1}\,\mathrm{d}x,
    \vartheta_{2}=p_{2}\,\mathrm{d}x$, respectively.
Within the overlap of \(U_1\) and \(U_2\), the two one-forms satisfy, respectively,
\begin{equation}
    \vartheta_2-\vartheta_1
    =:\mathrm{d}\Lambda_{12}=
    \begin{cases}
        0, & (p_2=p_1,
          2\pi-\delta<\operatorname{Re}p_2<2\pi)\\[1mm]
        2\pi\,\mathrm{d}x
        =\mathrm{d}(2\pi x), & (p_2=p_1+2\pi,
          2\pi<\operatorname{Re}p_2<2\pi+\delta).
    \end{cases}
\end{equation}
Accordingly, we define the transition function from \(U_1\) to \(U_2\)
as
\begin{equation}\label{trasition_fun}
    \Lambda_{12}(x)
    =
    \begin{cases}
        0,
        & (p_2=p_1,
          2\pi-\delta<\operatorname{Re}p_2<2\pi)\\[1mm]
        2\pi x,
        & (p_2=p_1+2\pi,
          2\pi<\operatorname{Re}p_2<2\pi+\delta)
    \end{cases},
\end{equation}
and the transition function from \(U_2\) to \(U_1\) can be expressed as $-\Lambda_{12}(x)$.}

{Using the above chart construction, we next describe a periodic orbit
\(O\), initialized at \((x(0),p(0))\) with
\(\operatorname{Re}p(0)=\delta\), that winds once around the momentum cylinder.
We decompose the orbit \(O\) into two nonoverlapping path segments, \(\Gamma_1\) and \(\Gamma_2\). 
The first segment \(\Gamma_1\), described
in \(U_1\), runs from \(\operatorname{Re}p_1=\delta\) to
\(\operatorname{Re}p_1=2\pi-\delta\). In the overlap region, we switch
to \(U_2\), where \(\Gamma_2\) continues to
\(\operatorname{Re}p_2=2\pi+\delta\). The trajectory is finally patched back to \(U_1\), where
$p_1=p(0)$ and $p_2=p_1+2\pi$ represent the same physical momentum, forming the periodic orbit.
Accordingly, the globally defined contour integral consists not only of the
integrals evaluated within the individual charts but also of the
transition terms associated with switching between them:
\begin{equation}\label{16}
    \mathcal{J}
    =
    \int_{\Gamma_1}p_1\,\mathrm{d}x
    +
    \int_{\Gamma_2}p_2\,\mathrm{d}x
    +
    \Lambda_{12}(x(0))\Big|_{p_2=p_1}
    -
    \Lambda_{12}(x(0))\Big|_{p_2=p_1+2\pi},
\end{equation}
where we have used the fact that, in the limit $\delta\to0$, the spatial coordinates at both chart-transition points approach $x(0)$.
According to Eq.~\eqref{trasition_fun}, the first transition term $\Lambda_{12}\left.\right|_{p_2=p_1}=0$ and the second term is equal to $\Lambda_{12}\left.\right|_{p_2=p_1+2\pi}=2\pi x(0)$.
As a result, Eq.\eqref{16} can be written as
\begin{equation}
    \mathcal{J}
    =\oint_O pdx
    -2\pi x(0)=-\oint_O xdp,
\end{equation}
where we use $\int_{\Gamma_1}p_1\,\mathrm{d}x
    +\int_{\Gamma_2}p_2\,\mathrm{d}x=\int^{x(0),p(0)+2\pi}_{x(0),p(0)}pdx=\oint_O pdx $ in the limit $\delta\rightarrow 0$.
It is noted that \(\oint_O p\,\mathrm{d}x\) denotes the integral accumulated
during one traversal of the orbit, for which
$x(T)=x(0),p(T)=p(0)+2\pi$.
It is evaluated after choosing a continuous branch of \(p\) rather than a globally defined contour integral on the momentum cylinder.
The discussion above demonstrates that, for Brillouin-zone-winding orbits, a global formulation of the coordinate-space contour integral on the cylindrical space requires the boundary term.
With this term included, the coordinate- and momentum-space formulations yield the same quantization condition, namely the natural momentum-space form in Eq.~\eqref{QCp}.}

\subsection{Symmetry of momentum-space semiclassical orbits and its constraint on the energy spectrum}\label{App.A4}
The $S_\eta$ symmetry of the classical Hamiltonian, Eq.~\eqref{seta}, imposes constraints on the dynamics and on the corresponding momentum-space semiclassical cycles. We first specify the notation: The coordinate and momentum operators transformed by $\eta$ are defined as
\begin{equation}\label{etaops}
    \hat{x}_\eta=\eta\hat{x}\eta^{-1},\qquad
    \hat{p}_\eta=\eta\hat{p}\eta^{-1},
\end{equation}
which satisfy
\begin{equation}
    \hat{x}_\eta|x_\eta\rangle=x_\eta|x_\eta\rangle,
    \qquad
    \hat{p}_\eta|p_\eta\rangle=p_\eta|p_\eta\rangle .
\end{equation}
Both $\hat{x}_\eta=\hat{x}_\eta(\hat{x},\hat{p})$ and $\hat{p}_\eta=\hat{p}_\eta(\hat{x},\hat{p})$ can be expressed as functions of $\hat{x}$ and $\hat{p}$. In the momentum-space path integral, the relevant classical symbols are defined by
\begin{equation}\label{xetasymbol}
    \frac{\langle x|\hat{x}_\eta(\hat{x},\hat{p})|p\rangle}{\langle x|p\rangle}=x_\eta(x,p),
    \qquad
    \frac{\langle x|\hat{p}_\eta(\hat{x},\hat{p})|p\rangle}{\langle x|p\rangle}=p_\eta(x,p).
\end{equation}
Thus the $\eta$ operation induces a map in complex phase space,
\begin{equation}\label{etamap}
    \eta:\xi\rightarrow \xi_\eta(\xi),
    \qquad
    \xi=\{x,p\}^{\mathsf T},\qquad
    \xi_\eta(\xi)=\{x_\eta(\xi),p_\eta(\xi)\}^{\mathsf T}.
\end{equation}
This classical map is defined by the quantum similarity transformation in Eq.~\eqref{etaops}. For example, if $\eta=\mathcal{P}$ is spatial inversion, then
\begin{equation}
    \hat{x}_{\mathcal{P}}=\mathcal{P}\hat{x}\mathcal{P}^{-1}=-\hat{x},
    \qquad
    \hat{p}_{\mathcal{P}}=\mathcal{P}\hat{p}\mathcal{P}^{-1}=-\hat{p},
\end{equation}
and Eq.~\eqref{xetasymbol} gives
\begin{equation}
    x_{\mathcal{P}}=-x,
    \qquad
    p_{\mathcal{P}}=-p .
\end{equation}

\begin{Theorem}\label{thm1}
If $\xi_1(t)=\xi(t)$ is a solution of Hamilton's canonical equations along the time contour $t$, then
\begin{equation}\label{orbitmap}
    \xi_2(t^*)=\xi_\eta[\xi^*(\mp t)]
\end{equation}
is also a solution, but along the time contour $t^*$. The sign $\mp$ corresponds to whether the $\eta$ operation involves a transpose or not.
\end{Theorem}

The proof relies on two ingredients: the semiclassical $\eta$ transformation is canonical, and the Hamiltonian satisfies the $S_\eta$ symmetry. Starting from
\begin{equation}
    [\hat{x},\hat{p}]=i,
\end{equation}
one obtains under the $\eta$ operation
\begin{equation}\label{commeta}
    [\hat{x}_\eta,\hat{p}_\eta]=\mp i,
\end{equation}
where the same sign convention is used throughout. In the $|p_\eta\rangle$ representation, the coordinate operator is represented as $\hat{x}_\eta=\mp i\partial_{p_\eta}$, and the plane wave satisfies
\begin{equation}
    \langle x_\eta|p_\eta\rangle=e^{\mp i p_\eta x_\eta}.
\end{equation}
The transformed momentum-space propagator can therefore be written as
\begin{equation}\label{Gpetaeta}
G(p_{\eta f},p_{\eta i},t)
=\int_{p_{\eta i}}^{p_{\eta f}}\mathcal{D}[x_\eta]\mathcal{D}[p_\eta]
\exp\left\{i\int_0^t\left[\pm x_\eta\dot p_\eta
-H(x(x_\eta,p_\eta),p(x_\eta,p_\eta))\right]\dd t'\right\}.
\end{equation}
The saddle-point equations are
\begin{equation}\label{eometa}
    \dot{x}_\eta=\mp\frac{\partial H(x(x_\eta,p_\eta),p(x_\eta,p_\eta))}{\partial p_\eta},
    \qquad
    \dot{p}_\eta=\pm\frac{\partial H(x(x_\eta,p_\eta),p(x_\eta,p_\eta))}{\partial x_\eta}.
\end{equation}
Writing them in symplectic form gives
\begin{equation}\label{eometaJ}
    \frac{\dd \xi_\eta(t)}{\dd t}
    =\mp J\left[\nabla_{\xi_\eta}H(\xi(x_\eta,p_\eta))\right]^{\mathsf T},
    \qquad
    J=\begin{pmatrix}0&1\\-1&0\end{pmatrix}.
\end{equation}
Changing variables from $\xi_\eta$ to $\xi$ with the Jacobian matrix
\begin{equation}\label{Meta}
    M_\eta(\xi)=\frac{\partial \xi_\eta(\xi)}{\partial \xi}
    =\begin{pmatrix}
    \partial_x x_\eta & \partial_p x_\eta\\
    \partial_x p_\eta & \partial_p p_\eta
    \end{pmatrix},
\end{equation}
we find
\begin{equation}
    \frac{\dd \xi(t)}{\dd t}
    =\mp M_\eta^{-1}(\xi)J\left[M_\eta^{-1}(\xi)\right]^{\mathsf T}
    \left[\nabla_\xi H(\xi)\right]^{\mathsf T}.
\end{equation}
Comparing this equation with the direct canonical equation
\begin{equation}\label{eomxi}
    \frac{\dd \xi(t)}{\dd t}=J\left[\nabla_\xi H(\xi)\right]^{\mathsf T},
\end{equation}
we obtain the canonical condition
\begin{equation}\label{canonical}
    M_\eta(\xi)J M_\eta^{\mathsf T}(\xi)=\mp J .
\end{equation}

We now prove the symmetry of the semiclassical solutions. Taking the complex conjugate of Eq.~\eqref{eomxi} gives
\begin{equation}
    \frac{\dd \xi^*(t)}{\dd t^*}
    =J\left[\nabla_{\xi^*}H^*(\xi)\right]^{\mathsf T}.
\end{equation}
Changing variables from $\xi^*$ to $\xi_\eta$ through $\dd\xi_\eta(\xi^*)=M_\eta(\xi^*)\dd\xi^*$ yields
\begin{equation}
    M_\eta^{-1}(\xi^*)\frac{\dd \xi_\eta(\xi^*(t))}{\dd t^*}
    =J M_\eta^{\mathsf T}(\xi^*)
    \left[\nabla_{\xi_\eta(\xi^*)}H^*(\xi)\right]^{\mathsf T},
\end{equation}
and hence
\begin{equation}\label{motioneta}
    \frac{\dd \xi_\eta(\xi^*(t))}{\dd t^*}
    =M_\eta(\xi^*)J M_\eta^{\mathsf T}(\xi^*)
    \left[\nabla_{\xi_\eta(\xi^*)}H^*(\xi)\right]^{\mathsf T}.
\end{equation}
Using the $S_\eta$ condition $H(\xi_\eta(\xi))=H^*(\xi^*)$ and the canonical condition~\eqref{canonical}, Eq.~\eqref{motioneta} becomes
\begin{equation}
    \frac{\dd \xi_\eta(\xi^*(t))}{\dd(\mp t^*)}
    =J\left[\nabla_{\xi_\eta(\xi^*)}H(\xi_\eta(\xi^*))\right]^{\mathsf T}.
\end{equation}
After substituting $t\rightarrow\mp t$, this is the standard canonical equation along the contour $t^*$,
\begin{equation}
    \frac{\dd \xi_\eta[\xi^*(\mp t)]}{\dd t^*}
    =J\left[\nabla_{\xi_\eta(\xi^*)}H(\xi_\eta(\xi^*))\right]^{\mathsf T}.
\end{equation}
This proves Theorem~\ref{thm1}.

\begin{Theorem}\label{thm2}
For two $S_\eta$-related periodic orbits $O_{1}(t)$ and $O_{2}(t^*)$, the contour integrals satisfy
\begin{equation}\label{Ipconj}
    \oint_{O_{2}(t^*)}-x_2(t^*)\dd p_2(t^*)
    =\left(\oint_{O_{1}(t)}-x_1(t)\dd p_1(t)\right)^* .
\end{equation}
\end{Theorem}

The proof uses the fact that $O_{1}$ and $O_{2}$ form an $S_\eta$-symmetric pair. By Stokes' theorem,
\begin{equation}\label{stokesp}
    \oint_{O_{2}}(-x_2\dd p_2)
    =\int_{\Sigma_2}\dd p_2\wedge\dd x_2
    =\int_{\Sigma_2}\dd p_\eta(\xi^*)\wedge\dd x_\eta(\xi^*) ,
\end{equation}
where $\Sigma_2$ is the surface enclosed by $O_{2}$. The canonical condition~\eqref{canonical} implies
\begin{equation}
[x_\eta(\xi^*),p_\eta(\xi^*)]_{\xi^*}
=\frac{\partial x_\eta}{\partial x^*}\frac{\partial p_\eta}{\partial p^*}
-\frac{\partial x_\eta}{\partial p^*}\frac{\partial p_\eta}{\partial x^*}
=\mp1 .
\end{equation}
Thus the symplectic two-form transforms as
\begin{equation}\label{wedgep}
\begin{aligned}
\dd p_\eta(\xi^*)\wedge\dd x_\eta(\xi^*)
&=\det[M_\eta(\xi^*)]\dd p^*\wedge\dd x^* \\
&=[x_\eta(\xi^*),p_\eta(\xi^*)]_{\xi^*}\dd p^*\wedge\dd x^*
=\mp\dd p^*\wedge\dd x^* .
\end{aligned}
\end{equation}
The transformation from $\xi_\eta(\xi^*)$ to $\xi^*$ maps $O_{2}(t^*)$ to $O_{1}^*(\mp t)$. The orientation change of the surface enclosed by the orbit compensates the sign in Eq.~\eqref{wedgep}. Combining Eqs.~\eqref{stokesp} and~\eqref{wedgep}, we obtain
\begin{equation}\label{S62}
\begin{aligned}
\oint_{O_{2}(t^*)}[-x_2(t^*)\dd p_2(t^*)]
&=\mp\int_{O_{1,\mp}^*}\dd p^*\wedge\dd x^* \\
&=\mp\oint_{O_{1}^*(\mp t)}[-x^*(\mp t)\dd p^*(\mp t)] \\
&=\left(\oint_{O_{1}(t)}[-x_1(t)\dd p_1(t)]\right)^* .
\end{aligned}
\end{equation}
For Brillouin-zone-winding orbits, the above proof based on Stokes' theorem is
not directly applicable, because such orbits are noncontractible in the
periodic momentum space and do not bound a surface. An additional condition on $S_\eta$ is therefore required:
\begin{equation}
    S_\eta:\mp x_\eta(\xi)\, \dd p_\eta(\xi)=x\,\dd p.
\end{equation}
This condition directly ensures that the two momentum-space action integrals satisfy the relation in Eq.~\eqref{S62}, without requiring the orbit to enclose a surface.
The time-reversal transformation considered in the main text satisfies this condition: $-x_{\mathsf{T}}\dd p_{\mathsf{T}}=x\dd p$. Equation~\eqref{S62} therefore remains valid for the Brillouin-zone-winding trajectories considered here. 
This is the proof of \textbf{Theorem}~\ref{thm2}.

\begin{Theorem}\label{thm3}
The energies and periods of two symmetric orbits, $O_{1}(t)$ and $O_{2}(t^*)$, are complex conjugates:
\begin{equation}
    E_1=E_2^*,
    \qquad
    T_1(E_1)=T_2^*(E_2).
\end{equation}
\end{Theorem}

Because the Hamiltonian is time independent, the energy is conserved along each orbit. If $\eta$ does not involve a transpose, we substitute $\xi_1(t)=\xi(t)$ and $\xi_2(t^*)=\xi_\eta(\xi^*(t))$ into the Hamiltonian and obtain
\begin{equation}
    E_1=H(\xi(t)),
    \qquad
    E_2=H(\xi_\eta(\xi^*(t))).
\end{equation}
Using $H(\xi_\eta(\xi^*))=H^*(\xi)$ gives $E_1=E_2^*$. If $\eta$ involves a transpose, we compare $\xi_1(t)$ and $\xi_2(-t^*)$ instead, and the same conclusion follows.

For the periods, we use the momentum-space action integral
\begin{equation}\label{periodp}
    T_1(E_1)=\frac{\partial}{\partial E_1}
    \oint_{O_{1}(t)}[-x_1\dd p_1].
\end{equation}
Taking the complex conjugate and applying Theorem~\ref{thm2},
\begin{equation}
\begin{aligned}
T_1^*(E_1)
&=\left(\frac{\partial}{\partial E_1}\oint_{O_{1}}-x_1\dd p_1\right)^* \\
&=\frac{\partial}{\partial E_1^*}
\left(\oint_{O_{1}}-x_1\dd p_1\right)^*
=\frac{\partial}{\partial E_2}
\oint_{O_{2}}-x_2\dd p_2
=T_2(E_2).
\end{aligned}
\end{equation}
This is consistent with the fact that the two time contours are complex conjugates. Theorem~\ref{thm3} is proved.

\begin{Theorem}\label{thm4}
If $E_n$ is an eigenenergy obtained from the momentum-space quantization condition, then $E_n^*$ is also an eigenenergy.
\end{Theorem}

For a quantized orbit $O_{n1}$, Eq.~\eqref{quantp} gives
\begin{equation}\label{q1}
    \oint_{O_{n1}(t)}-x_1(t)\dd p_1(t)=2\pi(n+\mu).
\end{equation}
The symmetric counterpart is $O_{n2}(t^*)$, whose energy is $E_{n2}=E_{n1}^*$ according to Theorem~\ref{thm3}. Theorem~\ref{thm2} ensures that
\begin{equation}\label{q2}
    \oint_{O_{n2}(t^*)}-x_2(t^*)\dd p_2(t^*)
    =\left(\oint_{O_{n1}(t)}-x_1(t)\dd p_1(t)\right)^*
    =2\pi(n+\mu).
\end{equation}
Therefore $E_{n2}=E_n^*$ is also selected by the same quantization condition.

\begin{Theorem}\label{thm5}
The pseudo-Hermitian spectrum has two possible semiclassical realizations. (1) $O_{1}$ and $O_{2}$ describe the same closed orbit with $S_\eta$ symmetry; the corresponding quantized eigenenergies satisfy $E_n=E_n^*$ and are real. (2) $O_{1}$ and $O_{2}$ are distinct orbits exchanged by $S_\eta$; the corresponding quantized eigenenergies form complex-conjugate pairs, $E_{n1}=E_{n2}^*$.
\end{Theorem}

In case (1), the two solutions describe the same closed orbit, denoted by $O$. They are solutions along the same time contour, so $t=t^*$ and the contour lies on the real-time axis. The two parametrizations may differ by a time shift $\Delta t$,
\begin{equation}
    O_{1}(t)=O_{2}(t+\Delta t).
\end{equation}
Combining this with Theorem~\ref{thm1} gives the $S_\eta$ symmetry of the orbit,
\begin{equation}\label{selfsymorbit}
    O(t-\Delta t)=\xi_\eta[O^*(\mp t)] .
\end{equation}
Since the two trajectories are the same orbit,
\begin{equation}
    E_1=H(O_{1}(t))=H(O_{2}(t+\Delta t))=E_2 .
\end{equation}
Together with \textbf{Theorem}~\ref{thm3}, this gives $E_1=E_1^*$; hence the quantized levels on such self-symmetric orbits are real.

In case (2), $O_{1}$ and $O_{2}$ are two different orbits related by the $S_\eta$ operation. \textbf{Theorem}~\ref{thm4} then implies that the quantized eigenenergies occur in complex-conjugate pairs.

\setcounter{equation}{0}
\renewcommand{\theequation}{B\arabic{equation}}
\renewcommand{\theHequation}{appendix.B.\arabic{equation}}
\setcounter{equation}{0}
\setcounter{figure}{0}
\setcounter{table}{0}

\renewcommand{\theequation}{B\arabic{equation}}
\renewcommand{\thefigure}{B\arabic{figure}}
\renewcommand{\thetable}{B\arabic{table}}

\section[\appendixname~\thesection]{Boundary-induced real-to-complex transition of Bloch oscillations}\label{App.B}
In this section, we investigate how boundary reflection modifies Bloch oscillations and the associated quantized spectrum. 
In the following, we focus on reflection from the left boundary; the effect of the right boundary can be analyzed analogously.
To isolate the effect of the left boundary, we consider a semi-infinite lattice with a boundary at (x=0) that extends to infinity on the right. This approximation is justified for a sufficiently long finite lattice, since states near the left boundary are insensitive to the distant right boundary, whose contribution can be ignored. Our analysis proceeds in two steps. 
We first derive the exact Green's function of the lattice model and then take its semiclassical limit, thereby resolving it into contributions from complete Bloch orbits and boundary-reflected orbits. 
The poles of the resulting semiclassical Green's function yield a generalized quantization condition containing both types of orbits. 
Next, we analyze the relative contributions of these orbits in distinct energy regimes and show that, when one orbit dominates, the generalized quantization condition reduces to the corresponding single-orbit form.
 

We first calculate the Green's function of the lattice version of the Hamiltonian in Eq.~\eqref{Hbloch}:
\begin{equation}
    \mathcal{H}=\sum_{m=1}^{\infty}
    \left(
    t_0 C_{m+1}^{\dagger}C_m
    -t_0 C_m^{\dagger}C_{m+1}
    +Fm C_m^{\dagger}C_m
    \right),
\end{equation}
where $C_m^\dagger$ and $C_m$ denote the creation and annihilation operators
at the $m$-th lattice site, respectively.
Here, we define the Green's function
$\hat{G}(E)=(E-\mathcal{H})^{-1}$ and denote its matrix elements in the
lattice-site basis by
$G(m,n',E)=\langle m|\hat{G}(E)|n'\rangle$.
To determine the Green's function, we start from the resolvent identity
\begin{equation}
    (E-\mathcal{H})\hat{G}(E)=\mathcal{I}.
\end{equation}
Taking its matrix elements in the lattice-site basis gives
\begin{equation}\label{connection}
    (E-Fm)G(m,n',E)-t_0G(m-1,n',E)+t_0G(m+1,n',E)=\delta_{m,n'}.
\end{equation}
This is a second-order inhomogeneous difference equation with respect to the lattice index $m$.
For $m\neq n'$, the source term vanishes, and the corresponding homogeneous equation admits two linearly independent solutions,
\begin{equation}
\begin{aligned}
    g^1_{m}(E)&=(-1)^m I_{m-E/F}\left(\frac{2t_0}{F}\right),\\
    g^2_{m}(E)&=K_{m-E/F}\left(\frac{2t_0}{F}\right),
\end{aligned}
\end{equation}
Here, $I_{\nu}(x)$ and $K_{\nu}(x)$ represent the modified Bessel functions of the first and second kind, respectively.
Using these two independent solutions, we construct the Green's
function separately in the regions $0<m<n'$ and $m>n'$, and then
match the two solutions at the source point $m=n'$.

In the region $0<m<n'$, the solution satisfying the left open-boundary
condition $G(0,n',E)=0$ can be written as
\begin{equation}
G(m,n',E)
=
c_{n'}(E)
\left[
g_0^2(E)g_m^1(E)
-
g_0^1(E)g_m^2(E)
\right].
\end{equation}

Because the Bloch oscillation is a local phenomenon,
the Green's function in the region $m>n'$ must decay as
$m\rightarrow\infty$. It can therefore be written as
\begin{equation}
G(m,n',E)
=
d_{n'}(E)g_m^1(E).
\end{equation}

The two solutions must coincide at $m=n'$ and satisfy the source
condition obtained from Eq.~\eqref{connection}, i.e.,
\begin{equation}
\begin{cases}
G(n',n',E)
=
c_{n'}(E)
\left[
g_0^2(E)g_{n'}^1(E)
-
g_0^1(E)g_{n'}^2(E)
\right]
=
d_{n'}(E)g_{n'}^1(E),
\\[3mm]
(E-Fn')G(n',n',E)
-
t_0c_{n'}(E)
\left[
g_0^2(E)g_{n'-1}^1(E)
-
g_0^1(E)g_{n'-1}^2(E)
\right]
+
t_0d_{n'}(E)g_{n'+1}^1(E)
=1.
\end{cases}
\end{equation}
Solving these equations gives the diagonal Green's function
\begin{equation}\label{G_exact}
\begin{aligned}
G(n',n',E)=\frac{2}{F}\Bigg[(-1)^{n'}
\frac{K_{E/F}\left(\frac{2t_0}{F}\right)
}{I_{-E/F}\left(\frac{2t_0}{F}\right)}I_{n'-E/F}^{\,2}\left(\frac{2t_0}{F}\right)
-I_{n'-E/F}\left(\frac{2t_0}{F}\right)K_{n'-E/F}\left(\frac{2t_0}{F}\right)\Bigg].
\end{aligned}
\end{equation}

We consider the parameter regimes
\begin{equation}
\left|\frac{Fn'-E}{F}\right|\gg1,
\qquad
\left|\frac{2t_0}{F}\right|\gg1,
\end{equation}
which corresponds to the Green's function to be evaluated away from the Bloch-oscillation center $n_c=E/F$ and the force $F$ is weak.
The approximation from Laplace's method gives the following asymptotic forms of the modified Bessel functions~\cite{bender2013advanced}:
\begin{equation}\label{Bessel_approx}
\begin{aligned}
K_{E/F}\left(\frac{2t_0}{F}\right)
&\simeq
\frac{\sqrt{\pi F/2}}
{\left(E^2+4t_0^2\right)^{1/4}}
\exp\left[\frac{W(0,E)}{2}\right],
\\[2mm]
I_{-E/F}\left(\frac{2t_0}{F}\right)
&\simeq
\frac{\sqrt{F/(2\pi)}}
{\left(E^2+4t_0^2\right)^{1/4}}
\left\{
\exp\left[-\frac{W(0,E)}{2}\right]
+
2\sin\left(\frac{\pi E}{F}\right)
\exp\left[\frac{W(0,E)}{2}\right]
\right\},
\\[2mm]
I_{n'-E/F}\left(\frac{2t_0}{F}\right)
&\simeq
\frac{\sqrt{F/(2\pi)}}
{\left[(Fn'-E)^2+4t_0^2\right]^{1/4}}
\exp\left[-\frac{W(n',E)}{2}\right],
\\[2mm]
K_{n'-E/F}\left(\frac{2t_0}{F}\right)
&\simeq
\frac{\sqrt{\pi F/2}}
{\left[(Fn'-E)^2+4t_0^2\right]^{1/4}}
\exp\left[\frac{W(n',E)}{2}\right],
\end{aligned}
\end{equation}
where
\begin{equation}
W(n',E)
=
\frac{2}{F}
\left[
(Fn'-E)
\operatorname{arsinh}
\left(
\frac{Fn'-E}{2t_0}
\right)
-
\sqrt{(Fn'-E)^2+4t_0^2}
\right].
\end{equation}
 
Inserting Eq.~\eqref{Bessel_approx} into Eq.~\eqref{G_exact}, we obtain the semiclassical Green's function
\begin{equation}\label{classical}
G(n',n',E)
\simeq
\frac{1}{\Delta_{n'}(E)}
\left[
\frac{
(-1)^{n'}e^{-W(n',E)}
}{
2\sin\left(\frac{\pi E}{F}\right)
+
\exp\left\{
\frac{2}{F}
\left[
\sqrt{E^2+4t_0^2}
-
E\operatorname{arsinh}
\left(\frac{E}{2t_0}\right)
\right]
\right\}
}
-1
\right],
\end{equation}
where
$\Delta_{n'}(E)=\sqrt{(Fn'-E)^2+4t_0^2}$
arises from the fluctuations around the
semiclassical orbits.

\begin{figure}
    \centering
    \includegraphics[width=0.8\linewidth]{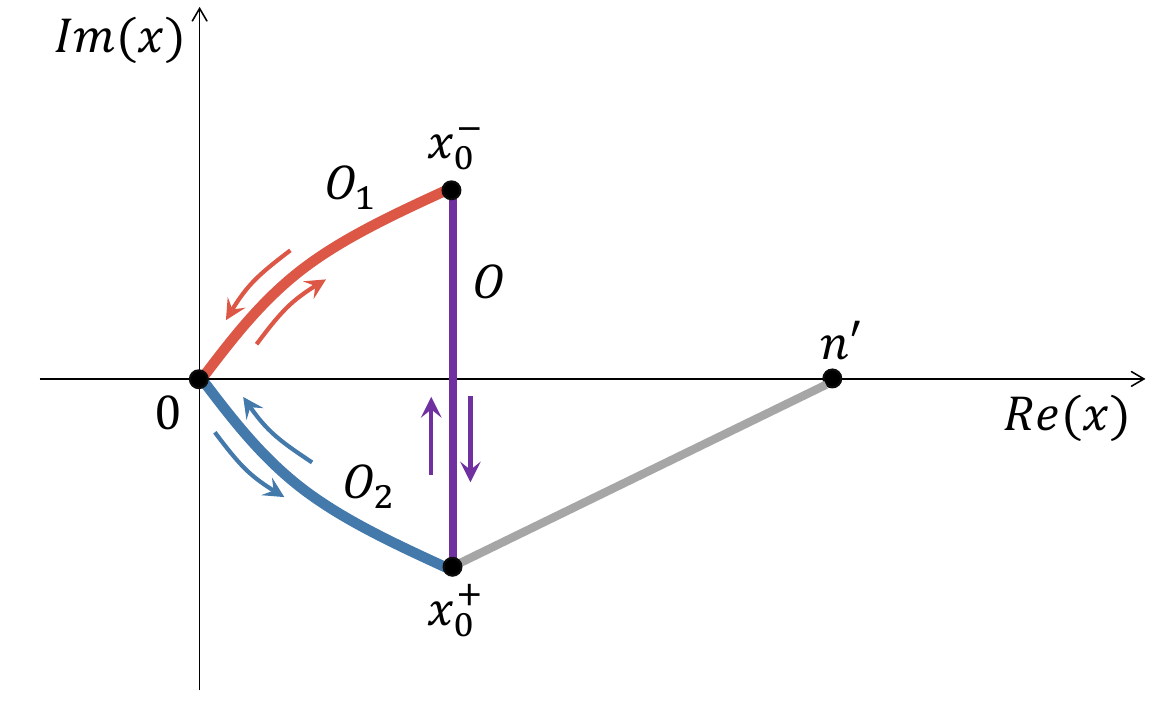}
    \caption{Schematic illustration of the orbits related to the semiclassical Green's function $G(n',n',E)$.}
    \label{fig:orbits}
\end{figure}

Next, we express the semiclassical Green's function in terms of the contributions from classical orbits. To this end, we first identify the relevant orbits. The classical Hamiltonian
\begin{equation}
    H(x,p)=2it_0\sin p+Fx
\end{equation}
possesses two turning points in complex phase space,
\begin{equation}
    x_0^{\pm}
    =
    \frac{E\mp2it_0}{F},
    \qquad
    p_0^{\pm}
    =
    \pm\frac{\pi}{2}.
\end{equation}
The complete Bloch orbit $O$ oscillates between these two turning points, as shown in Eq.~\eqref{motion}. In the presence of the left boundary at $x=0$, the two turning points further generate two boundary-reflected periodic orbits $O_1,O_2$. 
We choose the boundary point $x(0)=0$ as the initial point of the orbits.
The corresponding momenta at the boundary $x=0$ satisfy $E=2it_0\sin p_{\rm b}$,
the solutions of which are
\begin{equation}\label{pb}
    p_{\rm b}=-i\operatorname{arsinh}\left(\frac{E}{2t_0}\right),
    \qquad
    \widetilde{p}_{\rm b}=\pi-p_{\rm b}
    =\pi+i\operatorname{arsinh}\left(\frac{E}{2t_0}\right) .
\end{equation}
The two boundary-reflected orbits are parametrized by the following time evolutions,
\begin{equation}\label{gamma1}
O_1(t)=
\left\{
\begin{aligned}
    p_1(t)&=p_{\rm b}-Ft\\
    x_1(t)&=\frac{E-2it_0\sin\!\left(p_1(t)\right)}{F}\\
    t&:0\rightarrow T_1
\end{aligned}
\right.
\xrightarrow{\;\text{boundary reflection}\;}
\left\{
\begin{aligned}
    p_1(T_1)&=p_{\rm b}\\
    x_1(T_1)&=0\\
    t=&T_1
\end{aligned}
\right.,
\end{equation}
with the complex period $ T_1(E)=\frac{\pi-2i\operatorname{arsinh}\left(E/2t_0\right)}{F}$.
\(O_1\) describes an orbit that starts from $x=0$ with momentum $p_{\rm b}$, and evolves along a complex-time contour $0\rightarrow T_1$ and returns to the boundary with
momentum $p_1(T_1)=\widetilde p_{\rm b}-2\pi$ at the time $T_1$.
After the boundary reflection, the momentum returns to $p_{\rm b}$, thereby
forming the periodic orbit,
\begin{equation}\label{gamma2}
O_2(t)=
\left\{
\begin{aligned}
    p_2(t)&=\widetilde p_{\rm b}-Ft\\
    x_2(t)&=\frac{E-2it_0\sin\!\left(p_2(t)\right)}{F}\\
    t&:0\rightarrow T_2
\end{aligned}
\right.
\xrightarrow{\;\text{boundary reflection}\;}
\left\{
\begin{aligned}
    p_2(T_2)&=\widetilde p_{\rm b}\\
    x_2(T_2)&=0\\
    t=&T_2
\end{aligned}
\right.,
\end{equation}
with the period $T_2(E)=\frac{\pi+2i\operatorname{arsinh}\left(E/2t_0\right)}{F}=T_1^*(E^*)$.
\(O_2\) describes an orbit that starts from $x=0$ with momentum $\widetilde p_{\rm b}$, and evolves along a complex-time contour $0\rightarrow T_2$ and returns to the boundary with
momentum $p_2(T_2)= p_{\rm b}$ at the time $T_2$.
After the boundary reflection, the momentum returns to $\widetilde p_{\rm b}$, thereby
forming the periodic orbit.
The three relevant semiclassical orbits $O,O_1,O_2$ are schematically illustrated in Fig.~\ref{fig:orbits}.
The contour integrals of these three classical orbits are
\begin{equation}
\begin{aligned}
    \mathcal{J}_{B}&=-\oint_O x dp=\frac{2\pi E}{F}\\
    \mathcal{J}_1&=-\oint_{O_1} x\,\mathrm{d}p=\oint_{O_1} p\,\mathrm{d}x=\frac{\pi E}{F}+i\frac{4t_0}{F}
\left[\sqrt{1+\left(\frac{E}{2t_0}\right)^2}-\frac{E}{2t_0}\operatorname{arsinh}
\left(\frac{E}{2t_0}\right)\right],\\
\mathcal{J}_2&=-\oint_{O_2} x\,\mathrm{d}p=\oint_{O_2} p\,\mathrm{d}x=\frac{\pi E}{F}-i\frac{4t_0}{F}
\left[\sqrt{1+\left(\frac{E}{2t_0}\right)^2}-\frac{E}{2t_0}\operatorname{arsinh}
\left(\frac{E}{2t_0}\right)\right],
\end{aligned}
\end{equation}
where $\mathcal{J}_1+\mathcal{J}_2=\mathcal{J}_{B}$, only two of the three contour integrals are independent. It is therefore sufficient to retain two independent contour integrals in the semiclassical orbit expansion.
Introducing the integral from the site $n'$ to the turning point $x_0^+$ and back
\begin{equation}
 \mathcal{J}[n',x_0^+]=\int_{n'}^{x_0^+}p_{+}(x,E)\mathrm{d}x+\int_{x_0^+}^{n'} p_{-}(x,E)\mathrm{d}x=iW(n',E)-\pi n'+\frac{\pi E}{F},
\end{equation}
where $p_{+}(x;E)=\pi+i\operatorname{arsinh}[(E-Fx)/2t_0]$ and $p_{-}(x;E)=-i\operatorname{arsinh}((E-Fx)/2t_0)$,
the semiclassical Green's function Eq.~\eqref{classical} can be expressed in terms of \(\mathcal{J}_2(E)\), \(\mathcal{J}_{\mathrm B}(E)\) and \(\mathcal{J}[n',x_0^+]\):
\begin{equation}\label{semiG}
G(n',n',E)
\simeq
-\frac{i}{\Delta_{n'}(E)}
\frac{ e^{i\mathcal{J}[n',x_0^+]}}{
1-e^{i\mathcal{J}_{\mathrm B}(E)}-i e^{i\mathcal{J}_2(E)}
}-\frac{1}{\Delta_{n'}(E)}.
\end{equation}
Then it can be expanded as a sum over semiclassical orbits:
\begin{align}
G(n',n';E)
&\simeq
-\frac{i}{\Delta_{n'}(E)}
e^{i\mathcal{J}[n',x_0^+]}
\sum_{N=0}^{\infty}
\left[
e^{i\mathcal{J}_{\mathrm B}}
+i
e^{i\mathcal{J}_2}
\right]^N
-\frac{1}{\Delta_{n'}(E)}
\nonumber\\
=&
-\frac{i}{\Delta_{n'}(E)}
e^{i\mathcal{J}[n',x_0^+]}(1+e^{i\mathcal{J}_B}+ie^{i\mathcal{J}_2}+e^{i2\mathcal{J}_B}+(ie^{i\mathcal{J}_2})^2+2ie^{i(\mathcal{J}_B+\mathcal{J}_2)}+...)
-\frac{1}{\Delta_{n'}(E)}.
\end{align}
This expression makes the contributing semiclassical paths transparent.
The zeroth-order contribution corresponds to the propagation from
\(x=n'\) to the turning point \(x_0^{+}\) and back. 
Higher-order contributions arise from trajectories that first propagate from \(x=n'\) to \(x_0^+\), subsequently execute one or more complete Bloch-oscillation orbits, boundary-reflected orbits, or arbitrary concatenations of the two, and finally return to \(x=n'\). 
The term $-\frac{1}{\Delta_{n'}(E)}$ represents the local, zero-length contribution associated with propagation at the site \(n'\), which does not enter the quantization condition.
From the poles of Eq.\eqref{semiG}, we can obtain the generalized quantization condition:
\begin{equation}\label{general_QC}
    1-e^{i \mathcal{J}_B}-ie^{i\mathcal{J}_2}=0,
\end{equation}
or, equivalently, its explicit expression in terms of the energy
\begin{equation}\label{general_QC_E} 
\begin{aligned}
    1-e^{2\pi iE/F}-i e^{i\pi E/F}
\exp\left[-\frac{2}{F}\left(E\operatorname{arsinh}\frac{E}{2t_0}-\sqrt{E^2+4t_0^2}\right)\right]
&=0,\\
2\sin\left(\frac{\pi E}{F}\right)+
\exp\left[-\frac{2}{F}\left(E\operatorname{arsinh}\frac{E}{2t_0}-\sqrt{E^2+4t_0^2}\right)\right]
&=0.
\end{aligned}
\end{equation}
Next, we analyze the contributions of the two orbits to the generalized quantization condition and show how it reduces to the appropriate single-orbit quantization rules in different energy regimes.
We first consider the case of real energies. 
For sufficiently large $E$, the center of the Bloch orbit,
$n_c=E/F$, lies far from the left boundary. In this regime,
\begin{equation}
\left|
\exp\left[
-\frac{2}{F}
\left(
E\operatorname{arsinh}\frac{E}{2t_0}
-\sqrt{E^2+4t_0^2}
\right)
\right]
\right|
\ll 1,
\end{equation}
indicating that the contribution from the boundary-reflected orbit is exponentially suppressed. 
Neglecting this contribution, the
quantization condition reduces to
\begin{equation}
1-e^{i\mathcal{J}_B}=0,
\end{equation}
which is precisely the quantization condition for complete Bloch
oscillations.

As the energy decreases, the contribution from the boundary-reflected orbit gradually increases. 
For real energies, $2\sin(\pi E/F)$ is bounded within the interval $[-2,2]$. Therefore, once the magnitude of the boundary term $|e^{i\mathcal{J}_2}|$ in Eq.~\eqref{general_QC_E} exceeds $2$, i.e.,
\begin{equation}
\begin{aligned}
-\frac{2}{F}\left(E\operatorname{arsinh}\frac{E}{2t_0}-\sqrt{E^2+4t_0^2}
\right)&>\ln2,\\
   E<3.018t_0-0.2889&F,
\end{aligned}
\end{equation}
 the quantization equation no longer admits real-energy solutions.
 In the weak-field regime \(\lvert 2t_0/F\rvert\gg1\), this inequality reduces to $E<E_c^{l}=3.018t_0$, which corresponds to 
 \begin{equation}
     \operatorname{Im}\oint_{O_2}x\dd p>0,
 \end{equation}
meaning that the amplitude of $O_2$ exceeds that of $O$.
This shows that the real-to-complex spectral transition occurs at the critical point where the boundary-reflected orbit is no longer exponentially suppressed, becomes comparable to the complete Bloch-oscillation orbit, and subsequently emerges as the dominant semiclassical contribution.

Since the quantization condition admits no real-energy solutions in this
regime, $E$ must be analytically continued into the complex plane to
determine the corresponding eigenenergies. In this complex-energy regime,
the boundary-reflected orbit $O_2$ provides an essential contribution, and
the resulting quantization condition is determined by the relative
magnitude of the constant term and the complete Bloch-orbit contribution
$e^{i\mathcal{J}_{\mathrm B}}$.
We first consider the region with
\(\operatorname{Im}E>0\), where
\begin{equation}
\left|e^{i\mathcal{J}_{\mathrm B}}\right|
=\left|e^{2\pi iE/F}\right|
=
e^{-2\pi \operatorname{Im}E/F}
< 1.
\end{equation}
This indicates that the contribution from the complete Bloch-oscillation orbit is exponentially suppressed and can therefore be neglected. 
The quantization condition then reduces to
\begin{equation}
1-ie^{i\mathcal{J}_2(E)}=0,
\end{equation}
or equivalently,
\begin{equation}
\mathcal{J}_2(E)=2\pi\left(n+\frac{3}{4}\right).
\end{equation}
However, in the energy region with $\operatorname{Im}E<0$, where
\begin{equation}
\left|e^{i\mathcal{J}_{\mathrm B}}\right|
=
\left|e^{2\pi iE/F}\right|
=
e^{-2\pi\operatorname{Im}E/F}
> 1,
\end{equation}
the constant term in the general quantization condition is negligible compared with the two orbit contributions, and the condition
reduces to
\begin{equation}
\begin{aligned}
e^{i\mathcal{J}_{\mathrm B}}+ie^{i\mathcal{J}_2}&=0,\\
\Leftrightarrow \ \ \  1-ie^{i\mathcal{J}_1}&=0,
\end{aligned}
\end{equation}
where we have used $\mathcal{J}_{\mathrm B}=\mathcal{J}_1+\mathcal{J}_2$. Hence, the eigenenergies in the lower half of the complex-energy plane are semiclassically determined by the orbit \(O_1\) through
\begin{equation}
\mathcal{J}_1(E)=2\pi\left(n+\frac{3}{4}\right).
\end{equation}
Thus, we can divide the quantization condition into three regimes, depending on the values of $E$ in the complex-energy
plane:
\begin{equation}
\left\{
\begin{aligned}
-\oint_{O} x\,\mathrm{d}p
&=2\pi n,
&& E\in {R},\quad E>E_c^l,
\\[2mm]
\oint_{O_1} p\,\mathrm{d}x
&=2\pi\left(n+\frac{3}{4}\right),
&& \operatorname{Re}E<E_c^l,\quad \operatorname{Im}E<0,
\\[2mm]
\oint_{O_2} p\,\mathrm{d}x
&=2\pi\left(n+\frac{3}{4}\right),
&& \operatorname{Re}E<E_c^l,\quad \operatorname{Im}E>0.
\end{aligned}
\right.
\end{equation}
It should be noted that these single-orbit quantization conditions are
accurate in most of their respective energy regimes.
In the vicinity of $E^{l}_c$, the contributions from the different orbits become comparable, so that
neither of them can be neglected. 
In this case, the spectrum should be described by the general quantization condition
Eq.~\eqref{general_QC_E}.
{The exceptional point associated with the real-to-complex spectral transition lies in this multi-orbit regime, which is consistent with our previous conclusion that the exceptional point is inherently a quantum phenomenon, and it cannot be described by a single classical orbit.
\cite{cai2025prl}.}

\setcounter{equation}{0}
\renewcommand{\theequation}{C\arabic{equation}}
\renewcommand{\theHequation}{appendix.C.\arabic{equation}}
\setcounter{equation}{0}
\setcounter{figure}{0}
\setcounter{table}{0}

\renewcommand{\theequation}{C\arabic{equation}}
\renewcommand{\thefigure}{C\arabic{figure}}
\renewcommand{\thetable}{C\arabic{table}}
{
\section{Bloch Oscillations with Complex Next-Nearest-Neighbor Hopping}
\begin{figure}[ht]
    \centering
    \includegraphics[width=0.4\linewidth]{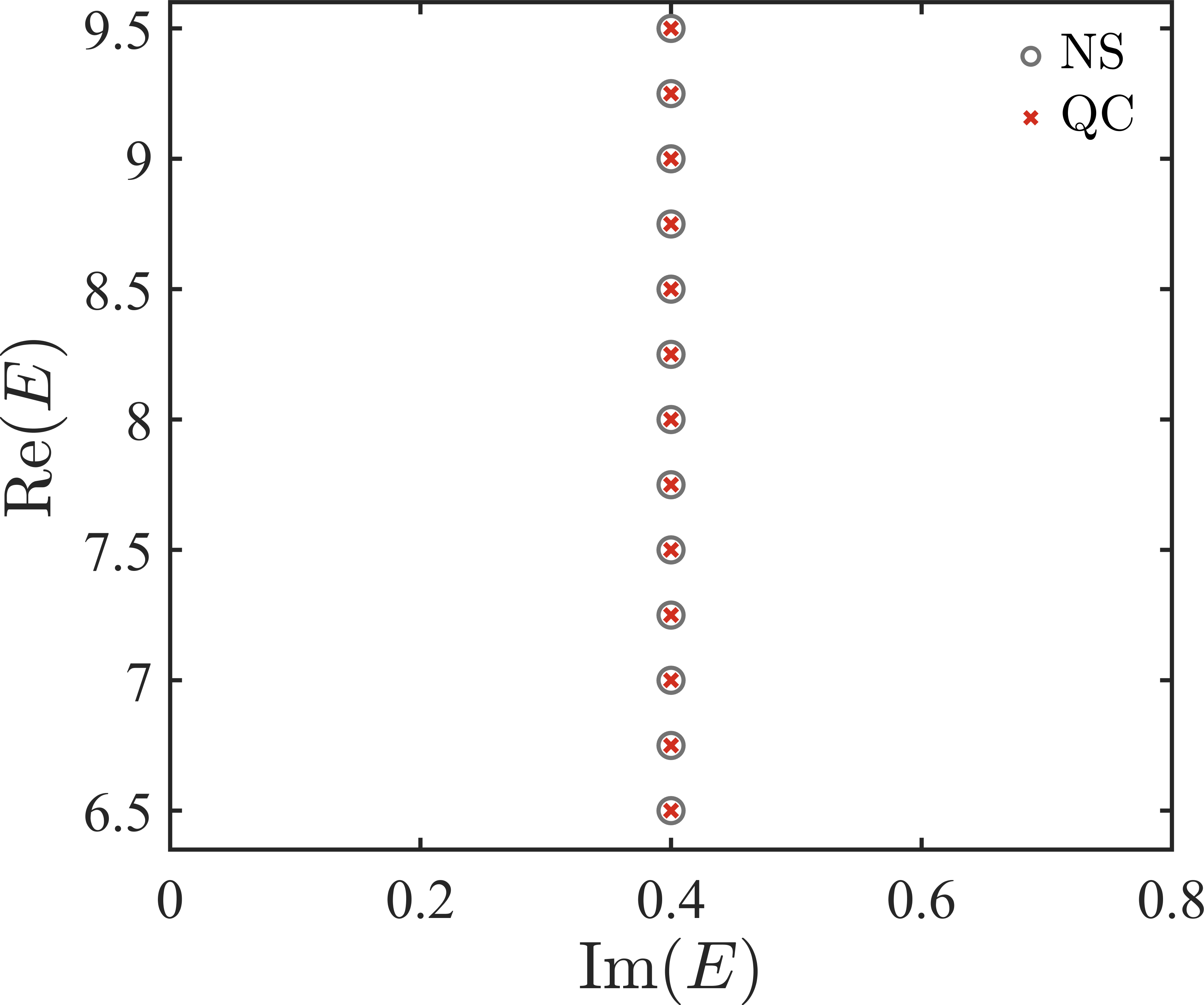}
    \caption{{Comparison between the bulk Bloch-oscillation energy levels obtained from numerical solution of the eigenvalue equation(NS, gray circles) and the analytical quantization condition (QC, red crosses) for the model with next-nearest-neighbor hopping. The parameters are \(F=0.25\), \(\gamma_0=0.4\), \(\gamma_1=1\), \(t_2=0.5\), and \(\gamma_2=0.3\).}}
    \label{fig:C1}
\end{figure}
In the main text, we derived the quantization condition in
Eq.~\eqref{21} and the corresponding energy levels in Eq.~\eqref{general_En} for Bloch
oscillations in a one-dimensional lattice with a periodic band dispersion
\(h(p)\). We then applied these results to the specific nearest-neighbor
hopping model in Eq.~\eqref{Hbloch}. To further verify the result in
Eq.~\eqref{general_En}, we now consider Bloch oscillations in a model with next-nearest-neighbor hopping, described by
\begin{equation}
    \mathcal{H}_2(\hat{x},\hat{p})
    =
    i\gamma_0
    +2i\gamma_1\sin\hat{p}
    +2t_2\cos(2\hat{p})
    +2i\gamma_2\sin(2\hat{p})
    +F\hat{x}.
\end{equation}
The corresponding lattice Hamiltonian is
\begin{equation}
\begin{aligned}
\mathcal{H}_2
=&
\sum_{m}
\left(i\gamma_0+Fm\right)C_m^{\dagger}C_m
+\gamma_1\sum_{m}
\left(
C_{m+1}^{\dagger}C_m
-
C_m^{\dagger}C_{m+1}
\right)
\\
&+t_2\sum_{m}
\left(
C_{m+2}^{\dagger}C_m
+
C_m^{\dagger}C_{m+2}
\right)
+\gamma_2\sum_{m}
\left(
C_{m+2}^{\dagger}C_m
-
C_m^{\dagger}C_{m+2}
\right).
\end{aligned}
\end{equation}
Using Eq.~\eqref{general_En}, we obtain the energy levels associated
with complete Bloch oscillations:
\begin{equation}\label{C3}
    E_n=i\gamma_0+Fn.
\end{equation}
As shown in Fig.~\ref{fig:C1}, the energy levels from the
quantization condition in Eq.~\eqref{C3} agree well with the numerical
results, providing further confirmation of the general result in
Eq.~\eqref{21}.
}



\begin{adjustwidth}{-\extralength}{0cm}

\reftitle{References}
\bibliography{pi_nh1}

\begin{thebibliography}{999}
\bibitem[con()]{contractibility}
{ The contractibility of a periodic orbit can be more formally
  characterized by its real-space-winding number $w_x=\frac{1}{L_x}\oint
  dx=\frac{x(T)-x(0)}{L_x}$ and its momentum-space-winding number
  $w_p=\frac{1}{2\pi}\oint dp=\frac{p(T)-p(0)}{2\pi}$, where \(L_x\) is the
  spatial period when the coordinate direction is periodic. When $x(T)=x(0)$
  and $p(T)=p(0)$, the vanishing of both \(w_x\) and \(w_p\) implies a
  contractible orbit, whereas a nonzero value of either $w_x$ or $w_p$
  indicates a noncontractible orbit.}
\end{thebibliography}

\PublishersNote{}
\end{adjustwidth}
\end{document}